\documentclass{aa}
\usepackage{graphics}

\newcommand{\gap}{\kern4.75pt}

\begin{document}

\title{The Westerbork HI Survey of Spiral and Irregular Galaxies\\ II.
R-Band Surface Photometry of Late-type Dwarf Galaxies\thanks{Based on
observations made with INT operated on the island of La Palma by the
Isaac Newton Group in the Spanish Observatorio del Roque de los
Muchachos of the Instituto de Astrofisica de Canarias.  The tables in
Appendix~A are only available in electronic form at the CDS via
anonymous ftp to cdsarc.u-strasbg.fr (130.79.128.5) or via
http://cdsweb.u-strasbg.fr/cgi-bin/qcat?J/A+A/ The figures in
Appendix~B are only available in electronic form
http://www.edpsciences.org}}

\author{R. A. Swaters\inst{1,2,3}\and M. Balcells\inst{4,1}}


\institute{Kapteyn Astronomical Institute, P.O.~Box 800, 9700 AV
  Groningen, The Netherlands\and Dept.\ of Physics and Astronomy,
  Johns Hopkins University, 3400 N. Charles Str., Baltimore, MD 21218,
  U.S.A. \and Space Telescope Science Institute, 3700 San Martin
  Drive, Baltimore, MD 21218, U.S.A.  \and Instituto de
  Astrof\'{\i}sica de Canarias, E-38200 La Laguna, Tenerife, Spain}

\date{Received date; accepted date}

\titlerunning{Surface photometry of late-type dwarf galaxies}
\authorrunning{R.A. Swaters \& M. Balcells}

\def\HI{\ifmmode \hbox{\scriptsize H\kern0.5pt{\footnotesize\sc i}}\else
H\kern1pt{\small I}\fi}
\def\HII{H\kern1pt{\small II}}
\def\figHI{H\kern1pt{\footnotesize I}}
\def\kms{km s$^{-1}$}
\def\etal{{et al}}
\def\varv{v}
\def\chapintro{1}
\def\chapopt{2}%
\def\chaphi{3}%
\def\chapkin{4}%
\def\chaplop{5}%
\def\chapdm{6}
\def\chapvdisp{7}
\def\chaplsb{8}%
\def\chapcon{9}
\def\chapnl{10}
\def\chapthanks{11}

\abstract{
$R$-band surface photometry is presented for 171 late-type dwarf and
irregular galaxies. For a subsample of 46 galaxies $B$-band
photometry is presented as well.  We present surface brightness
profiles as well as isophotal and photometric parameters including
magnitudes, diameters and central surface brightnesses.  Absolute
photometry is accurate to 0.1 mag or better for 77\% of the sample.
For over 85\% of the galaxies the radial surface brightness profiles
are consistent with published data within the measured photometric
uncertainty. For most of the galaxies in the sample \HI\ data have been obtained
with the Westerbork Synthesis Radio Telescope. The galaxies in our
sample are part of the WHISP project (Westerbork \HI\ Survey of Spiral
and Irregular Galaxies), which aims at mapping about 500 nearby
spiral and irregular galaxies in \HI. The availability of \HI\ data
makes this data set useful for a wide range of studies of the
structure, dark matter content and kinematics of late-type
dwarf galaxies.
}

\maketitle

\section{Introduction}
\label{introduction}

Over the years, a growing body of CCD-based data on the surface
brightness distributions of spiral and dwarf galaxies has become
available (e.g., de Jong \& van der Kruit 1994; Courteau 1996; Frei
\etal\ 1996; Heraudeau \& Simien 1996; Patterson \& Thuan 1996; Tully
\etal\ 1996; Matthews \& Gallagher 1997; Peletier \& Balcells 1997;
Jansen \etal\ 2000).  CCD imaging has been the impetus for many new
studies of a range of aspects of disk galaxies, such as general
scaling laws (de Jong 1996a; Courteau \etal\ 1996; Graham \& Prieto
1999; Jansen \etal\ 2000), the vertical light distribution (de Grijs
\& van der Kruit 1996; de Grijs 1998), the presence of thick disks
(Sackett \etal\ 1994; van Dokkum \etal\ 1994; de Grijs \& Peletier
1996), the radial truncation of the light distribution (Pohlen \etal\
2000; de Grijs \etal\ 2000), asymmetries and lopsidedness (Rix \&
Zaritsky 1995; Zaritsky \& Rix 1997; Kornreich \etal\ 1998; Conselice
\etal\ 2000), and the stellar populations of disk galaxies (de Jong
1996b; Peletier \& Balcells 1996; Bell \& de Jong 2000).

A combination of optical imaging data with HI observations, which
provide information both on the distribution of \HI\ and on the
kinematics of these galaxies, is a powerful tool to further our
understanding of the properties of disk galaxies, as has been
demonstrated by numerous papers in the literature in which one or a
few galaxies are discussed. To date, however, there does not exist in
the literature a large sample of galaxies for which both photometry
and \HI\ imaging exists.  The ongoing Westerbork \HI\ survey of spiral
and irregular galaxies (WHISP), which aims at mapping about 500 spiral
and irregular galaxies in \HI, is well suited for a more statistical
study of the link between optical and kinematical properties of disk
galaxies (for more details on the WHISP project and its goals, see
Swaters \etal\ 2000, hereafter Paper~I).

The two main aspects of the WHISP survey are a study of the \HI\ 
component of galaxies in itself, and a study of the kinematic
properties, focussing on rotation curves and dark matter properties.
As part of the WHISP survey, optical $R$-band images have been
obtained for the galaxies in the WHISP sample. These optical data
provide the deep optical images and accurate optical global
properties, such as disk surface brightnesses, disk scale lengths,
integrated magnitudes and optical diameters that are needed for a
detailed comparison of the optical, \HI\ and dark matter properties.
In addition to the two main aspects of WHISP, the combination of the
\HI\ distribution, the kinematics and the light distribution will
allow detailed studies of the nature of for example scaling laws
between \HI\ and optical properties, warps, truncated disks, and
lopsidedness.

In this paper, we present optical surface photometry data for the 171
late-type dwarf galaxies in the WHISP sample.  The outline of the
paper is as follows.  Sect.~\ref{sample} describes the selection of
the sample. In Sect.~\ref{thedist} the distance uncertainties for
the dwarf galaxies in our sample are discussed.
Sect.~\ref{observations} describes the observations and the data
reduction steps.  In Sect.~\ref{profiles} the ellipse fitting used
to derive the surface brightness, ellipticity and position angle
profiles is described.  Sect.~\ref{parameters} presents the global
photometric parameters obtained from the data, and
Sect.~\ref{comparison} describes the internal checks on our surface
photometry, the comparison of our profiles to those of other authors
and the comparison of global parameters to catalog values.
Sect.~\ref{discussion} gives a brief description of the optical
properties of the galaxies in this sample.  Finally,
Sect.~\ref{summary} gives a summary of the main results.

To facilitate the reading of text and tables, all long tables have
been placed at the end of the paper. Appendix~A presents the
tables with the selected sample, the assumed distances for all
galaxies, the list of observations and the derived optical properties.
Grayscale representations of each galaxy, together with surface
brightness profiles, are given in Appendix~B.

\section{Sample}
\label{sample}

The late-type dwarf galaxies presented here are part of the much
larger WHISP sample. This sample has been selected from the Uppsala
General Catalogue of Galaxies (UGC, Nilson 1973), and it consists of
all UGC galaxies with a blue major-axis diameter larger than $1.5'$,
$\delta(2000)>20^\circ$, and an \HI\ line flux density larger than 100
mJy (as calculated from the ratio of total \HI\ fluxes and profile
widths as listed in the Third Reference Catalogue of Bright Galaxies
(hereafter RC3, de Vaucouleurs \etal\ 1991)). These criteria ensure
sufficient resolution and a high enough signal-to-noise ratio for
observation with the Westerbork Synthesis Radio Telescope.

From the WHISP sample all late-type dwarf galaxies were selected for
the present study. The sample consists of two subsamples.  The first
consists of galaxies with \HI\ flux densities above 200 mJy, that
either have morphological types later than Sd, or that have earlier
morphological types and are fainter than $M_B=-17$.  This subsample
contains 113 late-type dwarfs and forms the basis of a study of \HI\
and dark matter in late-type dwarf galaxies (see Swaters 1999).
The second subsample comprises 80 galaxies with
flux densities between 100 and 200 mJy that were classified as dwarf
galaxies by Nilson (1973).  The total sample constructed in this way
contains 193 galaxies.  No upper limit was set for the diameter of the
selected galaxies.  For the 200 mJy sample, no lower limit to the
diameter was applied either, which resulted in the inclusion of four
dwarf galaxies with blue major-axis diameters smaller than $1.5'$.
Furthermore, there was no selection criterion based on the
environments of these dwarfs.  Isolated dwarfs as well as dwarf
companions to larger galaxies have been included in the sample.

Besides the selection effects of the UGC, which are well studied
(Thuan \& Seitzer 1979; Paturel \etal\ 1991; de Jong \& van der Kruit
1994), the present sample has an additional selection effect as a
result of the requirement that the galaxies have \HI\ measurements
listed in the RC3.  Not all galaxies have been observed in \HI\ (57\%
of the galaxies that meet all our criteria except the flux density
criterion have a measured \HI\ flux listed in the RC3), and those that
have been observed come from studies with different scientific goals,
different telescopes and different sensitivities.  In addition, the
selection based on the flux density may introduce a dependency on the
inclination and the kinematic properties of the galaxies.  Therefore
the true selection function is difficult to quantify.  With these
caveats in mind, the sample is representative of this galaxy
population as it spans the entire range of properties of late-type
dwarfs.

Because morphological type was one of the main selection criteria for
the sample selection, a few galaxies were included in the sample that
were assigned late morphological types, but that proved not to be
dwarf galaxies but large irregular galaxies, such as interacting
systems and peculiar galaxies.  Though these will be excluded in later
studies of the properties of late-type dwarfs, their optical
properties are presented here.

The sample of observed galaxies is listed in Table~\ref{sampletable}.
Out of the total of 193 galaxies, 171 have been observed.  The
remaining 22 have been missed due to bad weather. The galaxies in the
200 mJy sample were given priority, as a result only 2 out of the 113
galaxies in that sample have not been observed.

\section{Galaxy distances}
\label{thedist}

An important source of uncertainty for mass models of our dwarf sample
are the galaxy distances. Care has been given to using the best
distance estimator available for each galaxy.  Here we discuss the
choice of distance estimators and the magnitude of the distance
uncertainty associated to each.

The distances for the late-type dwarf galaxies in our sample have been
obtained from a search of the literature up to and including
1998. Most of the literature distances have been derived from four
distance indicators.  In order of decreasing priority these are based
on Cepheids, brightest stars, group membership and systemic
velocity. A full list of the adopted distances for all of the galaxies
in our sample, including the method used to determine the distance, is
given in Table~\ref{disttable}.  Unfortunately, Cepheid distances are
available for only three of the galaxies in our sample.  Below the
uncertainties for the three other methods are discussed.

\subsection{Brightest stars}

The accuracy of the brightest star method to determine the distance
has been subject of many discussions. There are a number of possible
problems, in particular if the brightest blue stars are used (e.g.,
Humphreys \& Aaronson 1987).  The brightest stars in a galaxy may not
be individual stars, but star clusters or compact \HII\ regions, or
these stars may be variable. In addition, it is found that the
luminosity of the brightest stars is dependent on the luminosity of
their parent galaxy, thus introducing a degeneracy in the distance
determination (e.g., Humphreys 1983). The brightest red stars suffer
much less from these problems.

The reported accuracy of the brightest stars as distance indicators
differs substantially between different studies. Rozanski \&
Rowan-Robinson (1994) found that the uncertainty in the distance modulus is
0.58 mag for the brightest red stars and 0.90 mag for the brightest
blue stars. Karachentsev \& Tikhonov (1994), on the other hand, find
uncertainties of 0.37 mag and 0.46 mag, respectively. Lyo \& Lee (1997)
found, from a comparison of distance determinations based on
Cepheids and brightest stars, that the uncertainties in the distance moduli
are 0.37 mag for the brightest red stars, and 0.55 for the brightest
blue stars, and they conclude that the brightest red stars are
therefore useful distance indicators.

Most of the distance determinations from brightest stars, listed in
Table~\ref{disttable}, are based on brightest red stars. With an
uncertainty in the distance modulus of 0.37 mag, the uncertainty in
the distance is about 20\%. If the uncertainty is as high as 0.58, as
suggested by Rozanski \& Rowan-Robinson (1994), the distance
uncertainty will be about 30\%.

\begin{figure*}[t]
\resizebox{\hsize}{!}{\includegraphics{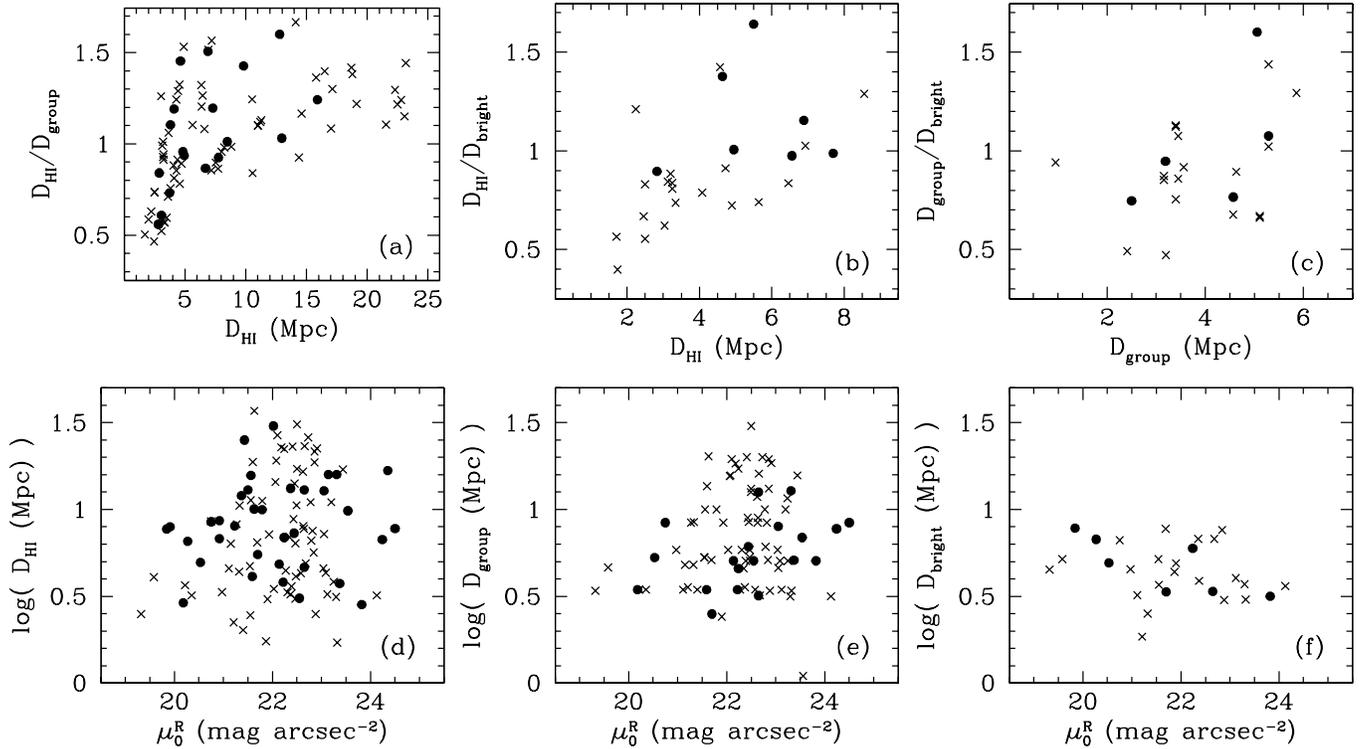}}
\caption{{\bf (a-c)} Ratios of distances derived from different
  distance indicators, plotted against the derived distances. The
  distances based on the \figHI\ systemic velocities are represented
  by $D_\mathrm{\HI}$, $D_\mathrm{group}$ refers to the group
  membership distances, and $D_\mathrm{bright}$ to the distances
  derived from the brightest stars. {\bf (d-f\kern1pt)} Derived
  distances versus $R$-band central disk surface brightness $\mu_0^R$.
  }
\label{figdist}
\end{figure*}

\subsection{Group membership}

Distances from group membership are mostly based on the groups
identified in de Vaucouleurs (1975) and de Vaucouleurs \etal\
(1983). The published distance moduli for these groups were
derived from several distance indicators: (a) from optical tertiary
indicators (morphological type and luminosity class); (b) from the
mean redshift of the group and a position-dependent Hubble constant
(varying between 70 \kms\ Mpc$^{-1}$ and 110 \kms\ Mpc$^{-1}$),
calibrated with spiral galaxies whose distances were determined from
tertiary distance indicators; (c) from the Tully-Fisher relation.
Unfortunately, de Vaucouleurs \etal\ (1983) do not list which
particular distance indicator or combination of distance indicators
was used to obtain the distance modulus they list. For the details on
these distance indicators, see de Vaucouleurs \etal\ (1983) and
references therein.

Given how the group membership distances have been determined, it is
clear that these distance may suffer from substantial uncertainties.
De Vaucouleurs (1979) claims that the uncertainty in the distance
moduli obtained from tertiary distance indicators is less than 0.4
mag. The Hubble constant based on tertiary distance indicators has an
average value of about 90 \kms\ Mpc$^{-1}$, higher than the currently
favored value of the Hubble constant, which may indicate that the
tertiary distance indicators tend to underestimate the distance.

Adding to this uncertainty is the fact that at small distances the
depth of the group may be a significant fraction of the group distance.

\subsection{Systemic velocity}

For all the galaxies in our sample a distance was calculated from the
\HI\ systemic velocity following the prescription given in
Kraan-Korteweg (1986) to correct for Virgocentric flow, with an
adopted Hubble constant of 75 \kms\ Mpc$^{-1}$. Many of the dwarf
galaxies have low systemic velocities, and these may well be dominated
by peculiar motions. As a result, these distances derived from the
systemic velocities may have large uncertainties, in particular for the
closest galaxies.

\subsection{Comparison of distances}

In the top panels of Fig.~\ref{figdist} the comparison between the
distances derived with the three different methods is shown.  In
Fig.~\ref{figdist}a the distances derived from the \HI\ systemic
velocities and group memberships are compared.  Two features stand
out. Firstly, at small distances there are a number of points that are
grouped along slanted lines.  These arise because there are a number
of large nearby groups to which a substantial number of dwarfs are
assigned, such as the M81 group, the M101 group and the NGC~4736
group. Secondly, at larger distances there appears to be a systematic
difference between the group distance and the distance derived from
the systematic velocity, in the sense that the group distances are
generally smaller. This may be the result of an underestimate of the
distances determined from tertiary distance indicators, as mentioned
above. On the other hand, almost all of these galaxies are in the same
region of the sky as the Virgo cluster. In this direction, the
relation between the systemic velocity and the distance is less
certain. This may also contribute to the differences between the two
distance estimates compared in Fig.~\ref{figdist}a.

Fig.~\ref{figdist}b shows a comparison of the distance from the \HI\
systemic velocity and the brightest star distance. There appears
to be a weak correlation between these two distance estimates. Towards
smaller \HI\ distances, the ratio of \HI\ distance over brightest star
distance decreases. All of the galaxies for which brightest star
distances are available are, because of the nature of the method, at
small distances.  At such small distances, the distances as derived
from the \HI\ systemic velocities are uncertain.

In Fig.~\ref{figdist}c the group distances are compared with the
brightest star distances. There is a slight indication that the
average ratio plotted in Fig.~\ref{figdist}c is below unity,
indicating that distances based on group membership are smaller than
those based on brightest stars, similar to what was concluded from
Fig.~\ref{figdist}a.

Another concern is that the distance estimates may depend on the
surface brightness of the galaxies. Although this is unlikely for the
distances derived from the \HI\ systemic velocities, it may play a
role for the group membership distances, which are partially based on
tertiary distance indicators, and for the brightest star distances. In
\hbox{Fig.~\ref{figdist}d-f} the different distance estimates are plotted
against central disk surface brightness $\mu_0^R$. There does not
appear to be a correlation between surface brightness and the \HI\
distance (Fig.~\ref{figdist}d), nor is such a correlation evident in
(Fig.~\ref{figdist}e, in which the group distance is plotted versus
surface brightness.

Finally, at best a weak trend is seen between the brightest star
distance and the surface brightness.  The variations from low to high
surface brightness are about 25\%, comparable to the random errors.
For low surface brightness galaxies smaller distances are found.
Because of the lack of a precise distance indicator, it is unclear
what causes this trend. It may be a result of selection effects. On
the other hand, it may also be that the magnitudes of the brightest
stars depend on surface brightness.

In summary, there are some indications that the group membership
distances may underestimate the true distances. This systematic
difference appears to be no larger than about 20\%. In addition, the
distance estimates based on brightest stars may depend on surface
brightness, but these effects are probably smaller than 25\%.  This
possible systematic error is likely to be dominated by the random
errors in the distance estimates. Based on Figs.~\ref{figdist}a-c, we
estimate that the largest uncertainty in distance is about 60\%, and
that the typical distance uncertainty is about 30\%.

\section{Observations and data reduction}
\label{observations}

As mentioned in the introduction, one of the usages of the optical
data is to provide the stellar distribution for mass modeling. For the
light distribution to be close to the stellar mass distribution,
observations at red wavelengths are necessary to minimize the effects
of extinction and young populations.  $I$-band observations were not
suitable because of the brighter sky in $I$, making longer exposures
necessary, and because of fringing making accurate flatfielding
difficult.  Even longer wavelengths, such as the $K$-band, were not
practical because the available infrared arrays had too small a field
of view for the galaxies in the sample.  This would make mosaicking
necessary and therefore require a lot of observing time.  The $R$-band
offered the best compromise. A possible problem in using $R$-band is
that the H$\alpha$ emission line is included in the wavelength range
covered by this filter.  Jansen (2000) has found that the median
contribution of H$\alpha$ to the $R$-band flux is 2.0\% for late-type
dwarf galaxies, with a maximum of 8\%.  Gallagher \& Hunter (1989)
show that radial H$\alpha$ profiles and the $R$-band radial surface
brightness profiles have similar shapes.  Hence, we expect that
the H$\alpha$ emission may introduce a systematic offset from the true
radial surface brightness profile of 0.02 mag on average, but the
shape of the profile will hardly be affected.

We used the f/3.29 prime focus camera at the 2.54m Isaac Newton
Telescope at La Palma, during six observing runs between May 1994 and
May 1996.  A list of the observing dates is given in
Table~\ref{runlist}.  In all runs, we used Harris filters. In the May
1994 run, we used an EEV CCD, in all other runs we used a thinned
Tektronix CCD (TEK3).  Specific details on the CCDs used are given in
Table~\ref{ccdchar}, which lists the readout noise, gain, pixel size,
chip size and field of view for each chip.  The EEV chip was always
used in standard readout mode, the TEK CCD was mostly used in quick
readout mode, saving time on readout of the chip.  The increased
readout noise was not important because the exposures are limited by
photon noise.  The emphasis was to reach faint surface brightness
levels.  For this reason, most galaxies were observed only in the
$R$-band.  However, for 46 galaxies we have obtained $B$-band data,
which are presented here as well.

Most galaxies were observed only once. Galaxies for which the center
was found to be saturated were observed a second time with a shorter
exposure. A list of all observations is given in Tables~\ref{obslistR}
and \ref{obslistB}. Some galaxies were observed in different observing
runs, and these data were used for internal comparison of the
photometric accuracy. Only the data with the best photometric
conditions are listed.  All of the data processing was done in {\sc
  iraf}.

\begin{table}
{\normalsize
\caption[]{List of observing runs}\label{runlist}
\begin{flushleft}
\begin{tabular}{l}
\hline\noalign{\smallskip}
May 1 -- May 6, 1994 (m94)\\
November 29 -- December 4, 1994 (d94)\\
February 2 -- February 6, 1995 (f95)\\
May 27 -- May 28, 1995 (m95)\\
December 23 -- December 28, 1995 (d95)\\
May 10 -- May 15, 1996 (m96)\\
\noalign{\smallskip}\hline
\end{tabular}
\end{flushleft}}
\end{table}

\begin{table*}
\caption[]{CCD characteristics\phantom{g}}\label{ccdchar}
\begin{flushleft}
\begin{tabular}{lcccccc}
\hline\noalign{\smallskip}
CCD & Readout & Gain & Noise (e) & Pixel size (\arcsec) & CCD size & Field
of
view \\
\noalign{\smallskip}\hline\noalign{\smallskip}
EEV & standard & 0.69 & 3.9 & 0.55 & 1280x1180 & $11.4'\times 10.5'$  \\
TEK & standard & 0.73 & 4.7 & 0.59 & 1124x1124 & $10.0'\times 10.0'$  \\
TEK & quick    & 1.47 & 6.2 & 0.59 & 1124x1124 & $10.0'\times 10.0'$  \\
\noalign{\smallskip}\hline
\end{tabular}
\end{flushleft}
\end{table*}

\subsection{Bias subtraction}

Both the EEV and the TEK3 had flat bias levels and low dark currents,
but the bias level did tend to vary slightly during the night.
Therefore, the bias level was determined from the overscan regions.
The May 1995 and the December 1995 run showed random bands in the bias
levels, which were visible throughout the image. The amplitude of
these bands was about 1 to 2 counts.  For these images, the bias level
was removed by subtracting from each image row the average of the
corresponding bias row.

\subsection{Flat fields}

The instrument was known to have light leaks at the filter wheel,
causing some excess light at the edges of the images.  By wrapping the
prime focus cone unit in a dark cloth, this effect was almost entirely
removed, allowing accurate flat fielding.

During all the runs, twilight flats were taken at the beginning and
the end of the nights, typically with 10000 to 30000
counts. Additionally, for increased accuracy night sky flats of blank
fields were taken, typically seven per filter (each exposure 180
seconds in R and 300 in B).

For each night and filter we constructed a set of flat fields, by
combining in different ways the twilight flats or the night sky flats.
When more than two flat fields were combined, a rejection algorithm
was used to remove stellar images and cosmic-ray events.  Otherwise,
stars and cosmic-ray events events were edited out by hand and replaced by
the local average.  Night sky flats were offset from one exposure to
the other, so that stars could be filtered out.  This filtering proved
to be satisfactory under good seeing conditions ($<1.5''$) with 6 or 7
exposures.  When the seeing was worse, or when fewer exposures had
been obtained, stellar residuals remained in the constructed flat
field.  In the latter case, flat fields were constructed by fitting a
low order two-dimensional polynomial to the night sky flat and to the
twilight flat; next, the twilight flat was divided by its fit, thus
obtaining the high signal-to-noise small scale variations, and these
were multiplied into the fit to the night sky flat.

After constructing these flat fields, a number of exposures of each
night was flat fielded by each of the constructed flatfields in order
to test which provided the most accurate results.  This was done by
checking the flatness of the empty regions. In more than half of the
cases the night sky flat fields allowed more accurate flat fielding
than the twilight flats.

Because of this careful construction and testing of the flatfields,
high flatfield accuracy was obtained.  For each exposure the sky
values at four places around each galaxy were measured, carefully
avoiding the outer parts of the galaxy and the halos of bright stars.
The background flatness was taken to be half the difference between
the minimum and maximum values. The flatness defined in this way has a
median value of 0.22\% for the sample in $R$, and 0.32\% in $B$.

\subsection{Calibration}

\begin{table}
{\normalsize
\caption[]{Calibration coefficients}\label{calcoeff}
\begin{flushleft}
\begin{tabular}{lrrrrr}
\hline\noalign{\smallskip}
Run & night & $c_{1,R}$ & $c_{2,R}$ & $c_{1,B}$ & $c_{2,B}$ \\
\hline\noalign{\smallskip}
May 94 & 1 & -0.480 & 0.196 & 0.697 & 0.223 \\
       & 2 & -0.489 & 0.160 &  ---  &  ---  \\
       & 3 & -0.510 & 0.184 &  ---  &  ---  \\
       & 4 & -0.462 & 0.149 & 0.564 & 0.287 \\
       & 5 & -0.448 & 0.145 &  ---  &  ---  \\
       & 6 & -0.591 & 0.285 &  ---  &  ---  \\
Dec 94 & 1 & -0.238 & 0.104 &  ---  &  ---  \\
       & 4 & -0.106 & 0.090 &  ---  &  ---  \\
       & 5 & -0.163 & 0.080 &  ---  &  ---  \\
       & 6 & -0.171 & 0.092 & 0.020 & 0.221 \\
Feb 95&1--5& -0.16  & 0.09  & -0.19 & 0.37  \\
May 95 & 1 &  0.116 & 0.035 & 0.361 & 0.175 \\
       & 2 &  0.102 & 0.064 & 0.389 & 0.175 \\
Dec 95 & 1 & -0.172 & 0.086 & 0.004 & 0.227 \\
       & 2 & -0.237 & 0.155 & 0.125 & 0.196 \\
    & 3--6 & -0.151 & 0.155 & 0.109 & 0.220 \\
May 96 & 1 & -0.061 & 0.104 & -0.076& 0.316 \\
       & 2 & -0.094 & 0.101 & -0.056& 0.263 \\
       & 4 & -0.091 & 0.123 & -0.104& 0.322 \\
\noalign{\smallskip}\hline
\end{tabular}
\end{flushleft}}
\end{table}

At regular intervals during each night, standard star fields from
Landolt (1992) were observed.  The fields we used were RU~149,
PG0231+051, PG0942-029, PG1323-086, PG1525-071, PG1528+062 and PG
2213-006. With these stars and the magnitudes given in Landolt (1992),
the observations were calibrated to Johnson $B$ and Kron-Cousins $R$.
The standard stars were observed over a range of airmasses for both
bands, both on photometric and non-photometric nights.  For all these
stars we determined the instrumental aperture magnitudes and spurious
values were deleted.  These data were combined to fit zero-point
magnitudes and color and extinction coefficients of the form:
\begin{equation}
\begin{array}{l}
b = B + c_{1,\,B} - 25 + c_{2,B}\,X + c_{3,B}\,(B-R)\\
r = R + c_{1,\,R} - 25 + c_{2,R}\,X + c_{3,R}\,(B-R)
\end{array}\label{photeq}
\end{equation}
where $B$ and $R$ are the standard star magnitudes, $b$ and $r$ are
the instrumental magnitudes, $X$ is the airmass, $c_1$ is the
zero-point offset, $c_2$ is the airmass term, and $c_3$ is the color
term. The color terms were found to be small, and can be neglected.
For most galaxies only $R$-band data are available, but the missing
color information does not significantly affect the accuracy of the
calibration.

The calibration was done for each night independently, except for a
few nights with poor photometric conditions, for which a combined
calibration was done. The derived calibration coefficients are listed
in Table~\ref{calcoeff}.  The $1\sigma$ residuals of the fit given by
Eq.~(\ref{photeq}) give the uncertainty on the calibration, and are
listed in Tables~\ref{obslistR} and \ref{obslistB} for the $R$ and
$B$-band, respectively.  This uncertainty also includes the
uncertainty introduced by ignoring the color term.  During nights with
cirrus, standard stars were observed as well and obviously, the
resulting photometric uncertainties are large.  For those galaxies for
which more than one observation was available, we used the one with
the best photometry.  If the seeing of the non-photometric
observations was better, or if that image had a higher signal-to-noise
ratio than the photometric one, we used the latter to calibrate the
non-photometric observation.

The majority of the sample has good photometric accuracy. The zero
point uncertainty is less than 0.1 mag for 77\% of the sample, and
below 0.2 mag for 89\% of the sample.

\subsection{Image combination}

All exposures of the same galaxy observed in the same run were aligned
using stars common in the frames.  If more than two exposures were
available, the images were combined using a rejection algorithm to
remove cosmic-ray events.  Otherwise, the cosmic-ray events were
removed using the {\sc figaro bclean} algorithm privately ported to
{\sc iraf} and replaced with the local average. If the central parts
of a galaxy were saturated in a long exposure, the saturated pixels
were rejected during the image combination and replaced with scaled
values from the short exposure images.

A few galaxies were too large to fit in the $10'\times 10'$ field of
view.  These galaxies were mosaicked.  All galaxies to be mosaicked were
observed in the December 1995 run, under poor photometric conditions.
The sky for all the images was subtracted and the images were aligned
using stars in overlapping regions.  Next, the images were scaled to the
same exposure time, and scaled photometrically to the frame with the
lowest sky and the highest counts per second, and the images
were combined.

\subsection{Final steps}

In each frame the mean value of the sky was determined in four areas
near the galaxy that were free of objects, stellar halos and scattered
light.  The sky value for each frame was the average of these four
mean values.  For the estimate of the uncertainty in the sky determination
we used half the difference between the minimum and the maximum of the
four mean values.  This estimate of the uncertainty in the sky includes
large scale variations in the flat fielding and also the uncertainty
in the sky determination itself.

For each exposure the seeing was estimated by fitting two-dimensional
Gaussians to all objects in the field that had been masked by hand
(see Sect.~\ref{profiles}). Only objects with ellipticities smaller
than 0.15 and central peaks more than five times the noise were used,
to avoid contamination by background galaxies and inaccurate
measurements. On average, about 40 objects were used to obtain the
seeing estimate. The seeing estimates listed in Tables~\ref{obslistR}
and \ref{obslistB} are the median values of the individual fits.
Given the pixel size of about $0.6''$, the observations with the best
seeing are undersampled. This affects approximately 15\% of the data
presented here.

The last step was to add the coordinate system to the frames.  The
coordinate system was transferred from the Digitized Sky Survey (DSS),
by determining coordinates from the plate solution for stars visible
in both the DSS and the CCD image.  Using these coordinates, a
coordinate system was fitted to the CCD frame, using about ten stars.
We used a four term matrix to describe the coordinate system in a
gnomonic projection.  The $1\sigma$ errors on the coordinate system
determined in this way are about $0.5''$.  This is comparable to the
typical error of $0.6''$ found by Veron-Cetty \& Veron (1996). The
$R$-band images of all galaxies, including the coordinate system, are
presented in Appendix~B.

\section{Isophotal fits}
\label{profiles}

Fitting ellipses to late type dwarf galaxies is not straightforward.
Often, the light distribution is irregular, and sometimes a well
defined center is missing, or the centers for the inner and outer
parts are different.  Nonetheless, the radial surface brightness
profile gives a useful representation of the light distribution.

Before fitting ellipses to the galaxy images, we masked out all the
stars and the scattered light in the frames by hand.  Special care was
taken to mask out low intensity halos of stars as well.  In some cases
this proved impossible, due to the proximity of a bright star to the
galaxy image.  In these cases, the foreground light was masked out to
where the galaxy light started to be dominant and the contribution of
the stellar halo was incorporated in the uncertainty in the sky
determination.

The ellipse fitting itself was done using the {\sc galphot} package
(Franx \etal\ 1989; J{\o}rgensen, Franx \&
Kj{\ae}gaard 1992).  The ellipse parameters were determined at
logarithmic intervals, each next ellipse having a radius 1.2 times
bigger than the previous ellipse, with the innermost radius at $1''$.
The radius $r$ of each ellipse is defined as $r=\sqrt{ab}$, where $a$
is the major axis and $b$ the minor axis. The ellipticity is defined
as $1-b/a$.  Determining the ellipse parameters was done in several
steps.  To make sure that the ellipse fits were not affected by
substructure, \HII\ regions, bars or other luminous components, all
fits were individually inspected at each step in the process, and
adjusted as described below when necessary.

First, we ran the fitting program leaving all the parameters (center,
position angle and ellipticity) free.  From these results we
determined the center.  If the galaxy had a well determined center,
e.g.\ a nuclear peak or a clear central condensation, this was adopted
to be its center.  In most other cases, we determined the center from
the average of the outer ellipses where the solution was more stable.
In those cases where the outer parts were highly irregular, the center
was determined from the average of the inner parts. The adopted center
for each galaxy is indicated in Appendix~B with a cross.

In the second step, the center was fixed.  From these results we
determined the position angle. This was done by averaging over the
outer ellipses, over the region where the results were converging.
The outermost points were discarded if these gave erratic
results.  If the position angle varied systematically with radius, the
average position angle was used.  For galaxies with a pronounced bar,
we tried to determine the position angle of the surrounding disk.

In the third iteration, the center and position angle were fixed, in
order to determine the ellipticity.  We determined the value for the
ellipticity in the same way as we determined the position angle.  The
ellipticities were set to zero if equal to zero within the
uncertainties of the measurements.

After fixing the orientation parameters, we extracted the calibrated
radial surface brightness profiles from all the images. The
orientation parameters as found for the $R$-band were also used to
extract the $B$-band radial surface brightness profiles.  In
Appendix~B we show the radial surface brightness profiles and the
exponential fits (see Sect.~\ref{parameters}) for all galaxies.
Also, the run of orientation parameters (center, ellipticity and
position angle) with radius are presented.  Note that these have been
derived from the free fits and may therefore, in some cases, deviate
substantially from the values chosen from subsequent fits with more
parameters fixed.

\section{Structural parameters}
\label{parameters}

\subsection{Profile fitting}

Most of the radial surface brightness profiles are smooth and regular,
despite the often irregular optical appearance of the galaxies.  Some
show a central excess of light, while others have a central
flattening.  In this paper, we have not attempted to make
decompositions into disk and central components. Instead, we focused
on the disk component in these galaxies.  An exponential intensity law
was fitted to the outer parts of the profiles, which becomes a linear
relation when expressed in magnitudes:
\begin{equation}
\mu(r)=\mu_0+1.0857\,{r\over{h}},\label{expdisk}
\end{equation}
where $\mu_0$ is the extrapolated central disk surface brightness, and
$h$ is the disk scale length.  The photometric errors were used as
weights.  Combined with the logarithmic spacing of the points in the
profile, this leads to higher weights to the central parts.  Most of
the profiles are well described by an exponential.  The profiles and
their fits are shown in Appendix~\ref{thefigures}. In a later paper,
we will study the properties of the galaxies with central light
concentrations with the use of the Sersic profile (Sersic 1968).

\subsection{Observed central surface brightness}

Beside the extrapolated central disk surface brightness $\mu_0$
derived from the fit, we also determined the observed central surface
brightness, $\mu_c$.  This value differs from $\mu_0$ if the profile
shows a central peak or flattening.  The observed central surface
brightness $\mu_c$ was determined from a linear extrapolation of the
surface brightness profile in the inner few arcseconds to $r=0$.  This
is justified given the exponential shape of the inner profile in most
cases, also for profiles with a central concentration of light.

\subsection{Diameters}

For each galaxy two sets of diameters have been determined.  Isophotal
diameters in both the $R$ and the $B$-band have been determined at
$\mu_R=25$ and 26.5 mag arcsec$^{-2}$.  Additionally, effective radii
within which 20\%, 50\% and 80\% of the light is contained have been
calculated.

\subsection{Extinction and inclination corrections}

The observed surface brightnesses and magnitudes suffer from
extinction, both from dust in our Galaxy and from that internal to the
observed galaxy.  The data were corrected for Galactic extinction using
the values for $A_B$ from Burstein \& Heiles (1984). The values for
$A_R$ have been obtained assuming an $A_B/A_R$ of 1.77 (Rieke \&
Lebofsky 1985).

\begin{figure}[t]
\resizebox{\hsize}{!}{\includegraphics{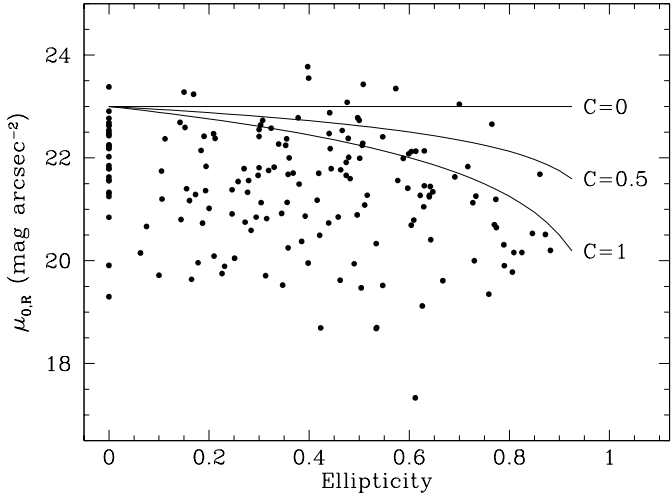}}
\caption{$R$-band central disk surface brightness
  $\mu_{0,R}$ versus ellipticity. The solid lines defined by
  Eq.~(\ref{extcor}). $C=0$ corresponds to optically thick, $C=1$ to
  optically thin.}
\label{intext}
\end{figure}

The correction for internal extinction is less certain.  A
relationship is expected between the axis ratio $b/a$ and the
observed surface brightness $\mu_0$ of approximately:
\begin{equation}
\mu_0 = \mu_0^i + 2.5\,C\log\,(b/a), \label{extcor}
\end{equation}
where $\mu_0^i$ is the surface brightness corrected for inclination,
and where $C$ is a constant related to the extinction. For the
transparent case, $C$ is equal to unity, for the opaque case it is
zero.

The dependence of $\mu_0$ on $b/a$ is shown in Fig.~\ref{intext}.
Note the relative paucity of high inclination, low surface brightness
galaxies. This might indicate that the galaxies in this sample are
transparent, and it would imply a lower cutoff in the true surface
brightness distribution of dwarf galaxies at $\mu_{0,R}=23$ mag
arcsec$^{-2}$. However, the lack of such galaxies may well be the
result of selection, e.g. as a result of the selection on flux
density.  In addition, given the selection on morphological type, some
galaxies may drop out of our sample if they may have been
misclassified as earlier morphological types when seen edge-on.  The
latter could contribute to the lack of high surface brightness edge-on
galaxies that are expected in the optically thin case.  In any case,
any expected trend is weak because of the wide range in surface
brightnesses at each $b/a$. Also, the correction is uncertain in the
cases where the ellipticity varies with radius. In conclusion,
Eq.~(\ref{extcor}) does not provide a useful way to determine the
transparency for this sample of galaxies.

\begin{figure}[t]
\resizebox{\hsize}{!}{\includegraphics{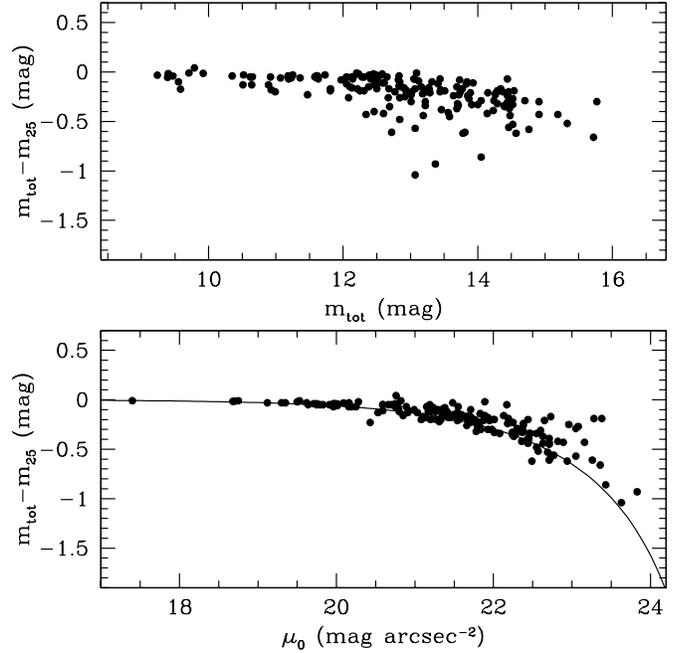}}
\caption{The extrapolation from $m_{25}$ to $m_\mathrm{tot}$ versus the
  integrated magnitude $m_\mathrm{tot}$ ({\it top panel}\kern1pt) and
  versus the extrapolated central disk surface brightness $\mu_0$
  ({\it bottom panel}\kern1pt). The solid line represents the
  correction for a purely exponential disk (see text).}
\label{extcomp}
\end{figure}

Because dwarf galaxies generally have low metalicities (e.g.,
Skillman \etal\ 1989), the dust content is likely to be
low, and therefore these galaxies are expected to have small internal
extinction. Therefore, $C=1$ was adopted, and in the remainder of this
paper the surface brightnesses were corrected accordingly. In the
optically thin case, the integrated magnitudes do not need to be
corrected. Note that the surface brightnesses and diameters listed in
Tables~\ref{globpropR} and~\ref{globpropB} have only been corrected
for Galactic extinction, not for inclination.

\subsection{Integrated magnitudes}

\begin{figure*}[pt]
\resizebox{\hsize}{!}{\includegraphics{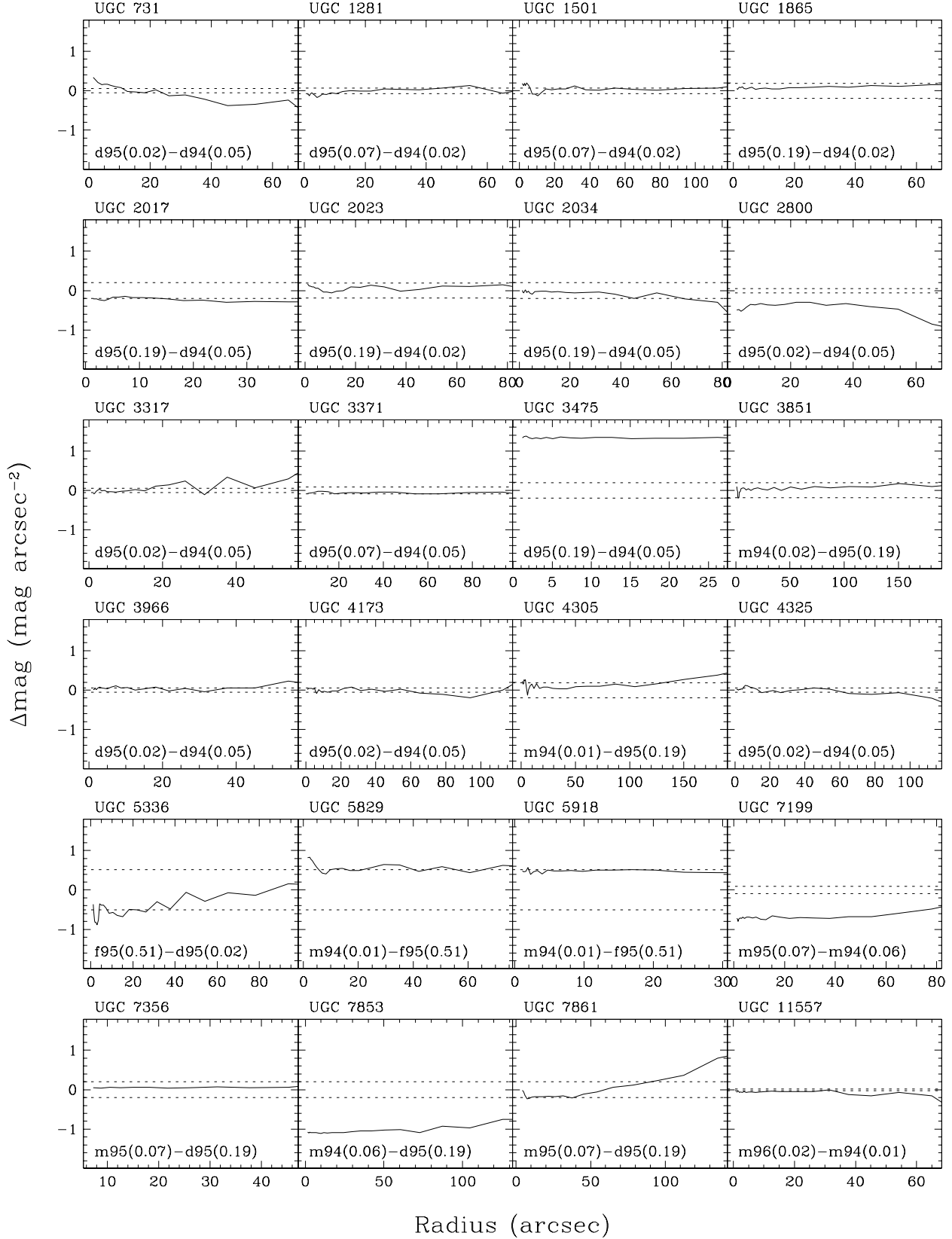}}
\caption{Internal comparison for the $R$-band observations. The radial
range for the comparison is set by the radius at which the profile
with the largest uncertainty reaches $3\sigma$ above sky. At the
bottom of each panel the observing runs that are compared are given,
the number between brackets gives the photometric accuracy. The dotted
lines in each panel give the combined $1\sigma$ errors. The shorthand
used to refer to the observing runs is explained in
Table~\ref{runlist}.}
\label{intcompR}
\end{figure*}

\begin{figure*}[t]
\resizebox{\hsize}{!}{\includegraphics{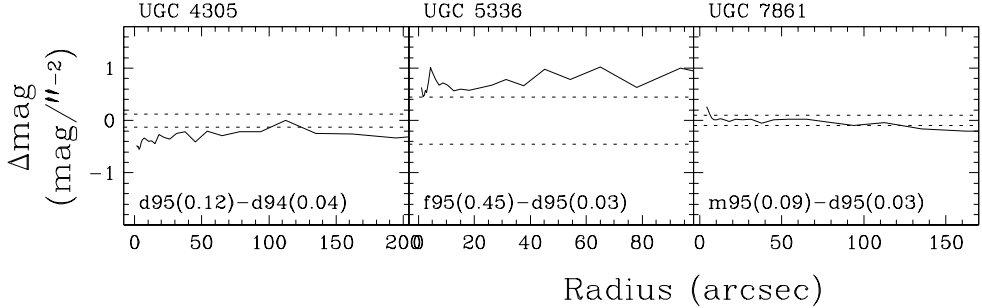}}
\caption{Same as Figure~\ref{intcompR}, but for the $B$-band observations.}
\label{intcompB}
\end{figure*}

Three magnitudes have been determined for each galaxy, two isophotal
($m_\mathrm{25}$ and $m_\mathrm{lim}$) and one total magnitude
($m_\mathrm{tot}$). The isophotal magnitudes are calculated at two
different levels. One at the 25 $R$ mag arcsec$^{-2}$, and one at the
limiting surface brightness level, which corresponds to the $3\sigma$
above sky level.  The total magnitude has been calculated by
extrapolating an exponential fitted to the outer parts of the surface
brightness profile out to infinity.  In particular for galaxies with
low surfaces brightnesses, such as the dwarfs in this sample, this
extrapolation can be significant.  To first order, the extrapolation
can be determined by assuming that the entire profile follows an
exponential decay, as was done by Tully et al.\ (1996).  However,
applying such a correction overestimates the total magnitude when a
central condensation is present, and underestimates it when the light
profile flattens towards the center, as Tully \etal\ (1996) already
noted.  To avoid this problem, we used the more general method to
determine the total magnitude (Han 1992).  This method assumes that
the light falls exponentially outside the last point in the profile,
but it makes no assumptions on the shape of the inner profile:
\begin{eqnarray}
m_\mathrm{tot} &=& m_\mathrm{lim} \nonumber \\
&&-2.5\log\left\{
1+{b\over{a}}\,
q\!\left({r_\mathrm{\,lim}\over h'}\right)
10^{-0.4(m_\mathrm{t}-m_\mathrm{lim})}
\right\},
\label{hancor}
\end{eqnarray}
where the function $q(x)$ reflects the luminosity-radius relationship
for such an exponential disk,
\begin{equation}
q(x) = (1+x)\,e^{-x},
\end{equation}
$r_\mathrm{lim}$ is the radius where $3\sigma$ is reached, and
\begin{equation}
m_\mathrm{t}=\mu_0'-5\log\, h' -2.5\log\, 2\pi
\label{msubt}
\end{equation}
is the total magnitude of a pure exponential disk characterized by
$\mu_0'$ and $h'$, as determined from fitting to the outer parts of
the profile.

The top panel of Fig.~\ref{extcomp} shows the extrapolation from
$m_{25}$ to $m_\mathrm{tot}$ versus the total magnitude
$m_\mathrm{tot}$. The corrections become larger for less luminous
galaxies, and may reach up to one magnitude. The bottom panel of
Fig.~\ref{extcomp} shows the extrapolation versus $\mu_0$. The
correction becomes larger for fainter surface brightnesses. The solid
line shows the extrapolation for a purely exponential disk. The
corrections generally follow this line closely.  However, some points
deviate up to about 0.5 mag from this line.  This plot shows that just
applying an extrapolation based on the assumption that the light
profile is exponential may give total magnitudes that are
significantly too bright, in particular for low surface brightness
galaxies.

\section{Comparison of radial surface brightness profiles}
\label{comparison}

In order to assess the photometric quality of our data and to check the
reliability of our estimate of the photometric uncertainty, radial surface
brightness profiles of identical galaxies obtained during different
runs have been compared. Our data have been compared to data available
in the literature as well.

\subsection{Internal comparison}

\begin{figure*}[pt]
\resizebox{\hsize}{!}{\includegraphics{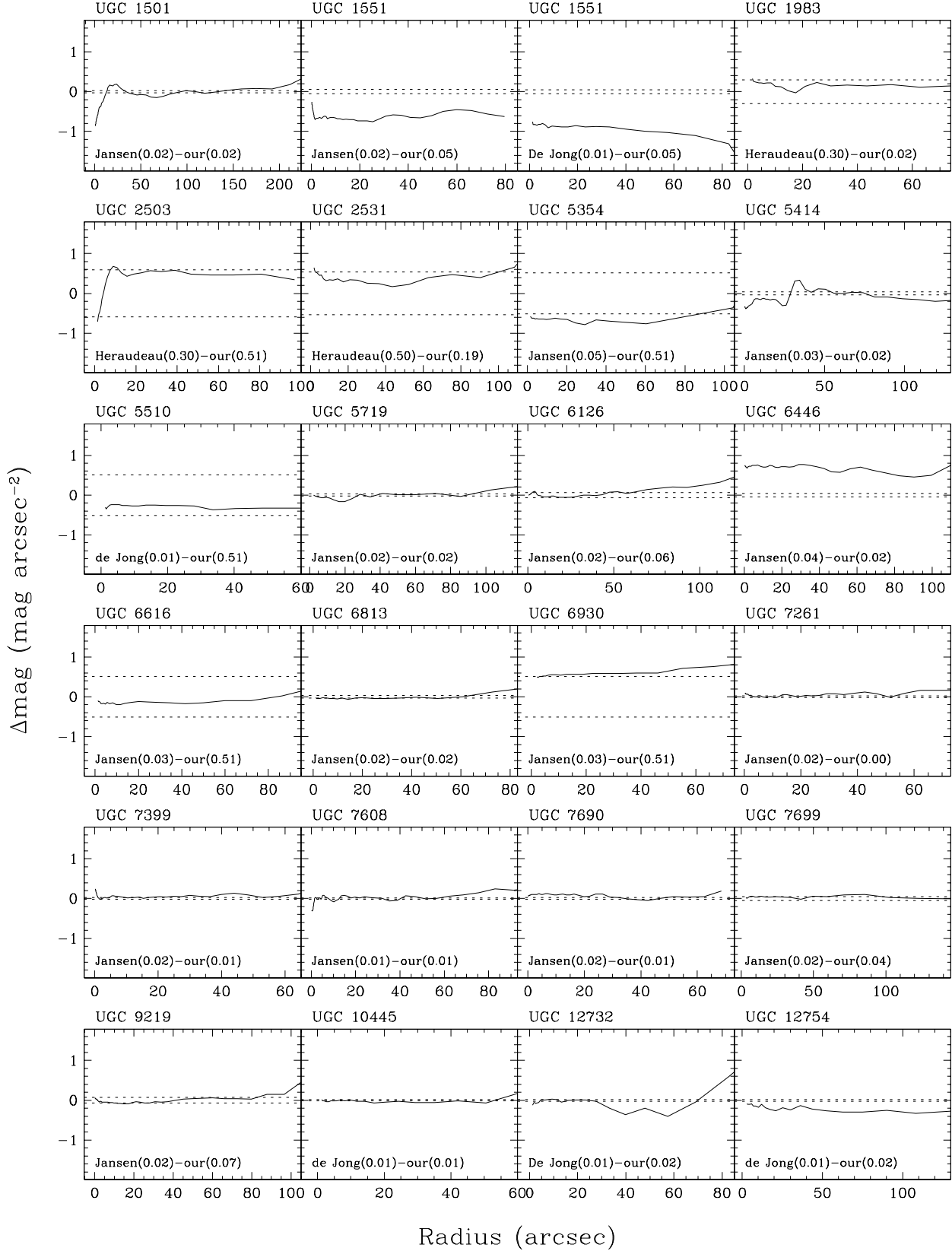}}
\caption{
  External comparison for $R$-band observations. The radial range for
  the comparison is set by the radius at which the profile with the
  largest uncertainty reaches $3\sigma$ above sky. At the bottom of
  each panel the data that are compared are given, the number between
  brackets gives the photometric accuracy. The dotted lines in each
  panel give the combined $1\sigma$ errors. In this figure we have
  included non-dwarf galaxies that were observed as part of the WHISP
  program and that will be published in a later paper. }
\label{compextR}
\end{figure*}

\begin{figure*}[t]
\resizebox{\hsize}{!}{\includegraphics{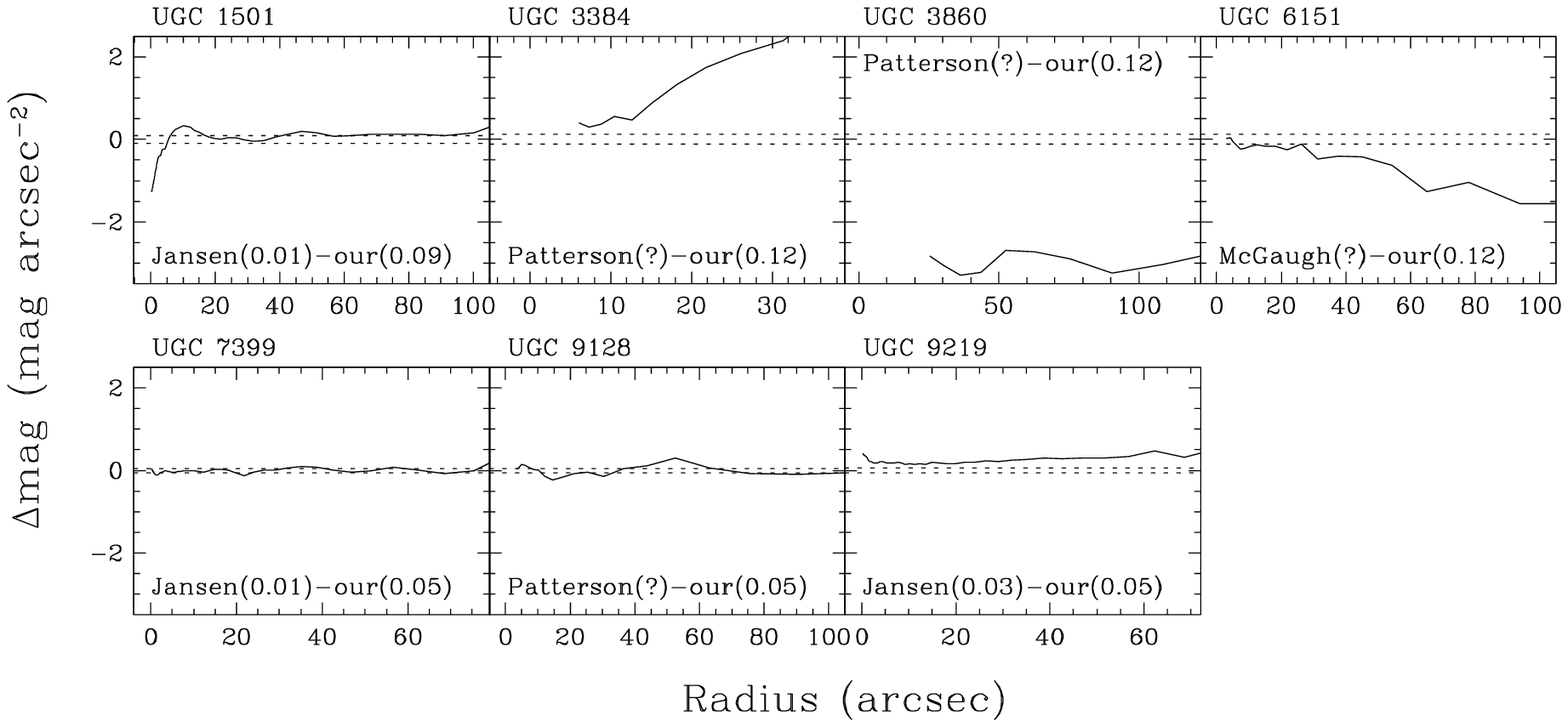}}
\null\vspace{-0.6cm}
\caption{Same as Figure~\ref{compextR}, but for the $B$-band observations.}
\label{compextB}
\vspace{0.5cm}
\end{figure*}

Some galaxies were observed on more than one night. Although for each
galaxy only one observation is given in this paper, these multiple
observations have been used to assess the quality of our calibration.
Figs.~\ref{intcompR} and~\ref{intcompB} show the internal comparison
for the $R$ and the $B$-band respectively. The dotted lines indicate
the $\pm1\sigma$ range based on the uncertainties on the photometric
solution. The majority of the profiles are equal within the
uncertainties of the measurements.  Five cases, \object{UGC~2800},
\object{UGC~3475}, \object{UGC~7199}, \object{UGC~7853} and
\object{UGC~7861} deviate more than expected from our estimates of the
uncertainty in the photometry.  These galaxies were all observed on
nights during which there was some cirrus present.  The uncertainty
caused by cirrus is in principle included in the photometric
uncertainty.  The deviant cases are probably the result of somewhat
thicker cirrus clouds passing overhead occasionally. Because in 83\%
of the internal comparisons the photometric uncertainty is accurately
known, including nights with some cirrus, we conclude that the
photometric uncertainty we have derived is a reliable estimate of the
real photometric uncertainty.  The comparison for the $B$-band
profiles, though available for only three galaxies, is consistent with
this conclusion.  Nights with cirrus have been marked in
Tables~\ref{globpropR} and~\ref{globpropB}.  The fact that the
difference profiles in Figs.~\ref{intcompR} and~\ref{intcompB} are
quite flat indicates that, for 80\% of the sample, uncertainties due
to flatfielding, sky brightness determination and contamination by
halos of foreground stars together are below 0.1 mag arcsec$^{-2}$
over the entire radial range.

\subsection{External comparison}

Comparing surface brightness profiles with those obtained by other
authors is not straightforward.  For the comparison with published
data, we restricted ourselves to CCD data.  We have used data by
McGaugh \& Bothun (1994), de Jong \& van de Kruit (1994), Heraudeau \&
Simien (1996), Patterson \& Thuan (1996) and Jansen \etal\ (2000).
Where the authors have derived surface brightness profiles with fixed
position angle and ellipticity, our profiles were rederived with the
same orientation parameters.  McGaugh \& Bothun (1994) and Patterson
\& Thuan (1996) used free fits, making comparison of the profiles
difficult.  To get a substantial sample for comparison with other
data, not only the data for the dwarf galaxies were used, but also the
data for other galaxies that were observed in this program.

\begin{figure}[t]
\resizebox{\hsize}{!}{\includegraphics{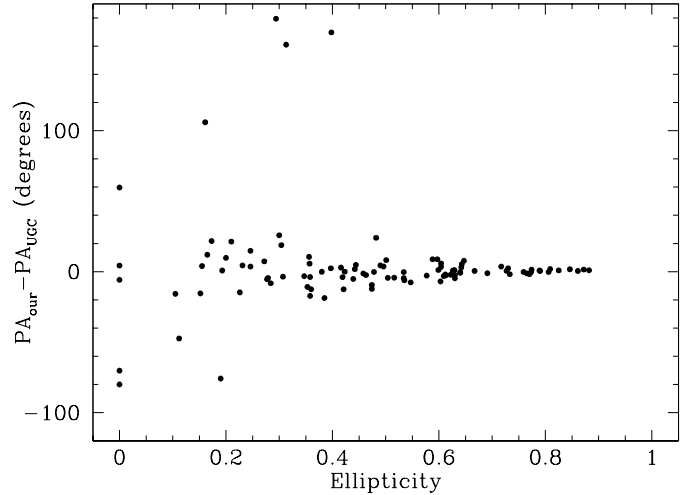}}
\caption{Comparison of our position angle and the UGC position angle
  as a function of ellipticity.}
\label{pacomp}
\end{figure}

Fig.~\ref{compextR} shows the comparison for the $R$-band profiles.
Most profiles are consistent within the $1\sigma$ errors, as indicated
by the dotted lines.  For three galaxies, our results deviate
significantly from the profiles obtained by others: \object{UGC~1551},
\object{UGC~6446} and \object{UGC~12754}.  For \object{UGC~1551} there
are two independent measurements available in the literature, by
Jansen \etal\ (2000) and de Jong \& van der Kruit (1994), which are in
agreement with each other.  The quality of those data for these
galaxies is high, as judged by the authors.  The three deviating
galaxies were observed during nights that suffered from occasionally
thicker cirrus.  Based on the external comparison, we conclude that
for 87\% of our observations the quoted photometric uncertainty is an
accurate estimate of the photometric quality, similar to the number
found from the internal comparison.

\begin{figure}[ht]
\resizebox{\hsize}{!}{\includegraphics{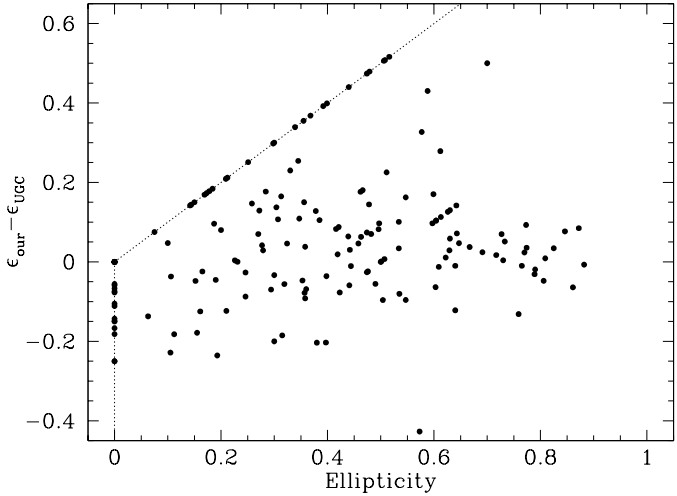}}
\caption{Comparison of our ellipticity and the UGC ellipticity as a
  function of our ellipticity. The dotted line corresponds to
  $\epsilon_{\mathrm our}=0$ or $\epsilon_{\mathrm UGC}=0$}
\label{ellcomp}
\vspace{1cm}
\end{figure}

The comparison with the published surface brightness profiles for the
$B$-band is presented in Fig.~\ref{compextB}. The comparison with the
data by Jansen \etal\ (2000) shows that our data is consistent within
the uncertainties with his. However, the comparison with the data by
McGaugh \& Bothun (1994) and Patterson \& Thuan (1996) shows clear
inconsistencies. Both groups of authors have used free fits rather
than fixed fits as we did. This explains the large difference found
for \object{UGC~3384}, because this galaxy is highly asymmetric.
Neither of the authors give uncertainties for their data. Except for
\object{UGC~3860}, the profiles are fairly consistent, in particular
in the inner parts. 

In conclusion, based on both the internal and external comparison, we
find that our quoted photometric uncertainties are accurate estimates of the
real photometric uncertainties.  In about 13\% of the cases the real
photometric uncertainties are underestimated. However, these poor nights are
overrepresented in this comparison because many of the galaxies
observed in non-photometric nights were deliberately reobserved in
order to obtain better photometry.  It is therefore likely that for a
larger fraction than the 87\% quoted above the photometric is uncertainty is
accurate.

\subsection{Comparison to UGC}

\begin{figure}[ht]
\resizebox{\hsize}{!}{\includegraphics{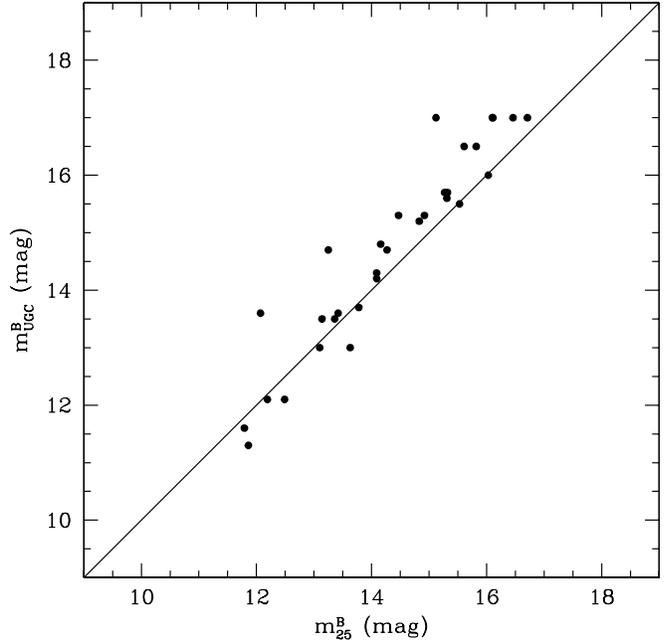}}
\caption{Comparison of the isophotal $B$-band magnitude $m_{25}^B$ and
  the magnitude quoted in the UGC. The solid line is the line of
  equality.}
\label{magcompb}
\vspace{0.2cm}
\end{figure}

An additional source to check our data against is the UGC.
Fig.~\ref{pacomp} shows the differences between our position angle and
those listed in the UGC. For highly inclined galaxies the agreement is
excellent. Towards more face-on galaxies the scatter increases and
some large differences occur. The outcome of this comparison in no
doubt affected by the fact that we have determined the position angles
at fainter surface brightness levels.

\begin{figure}[ht]
\resizebox{\hsize}{!}{\includegraphics{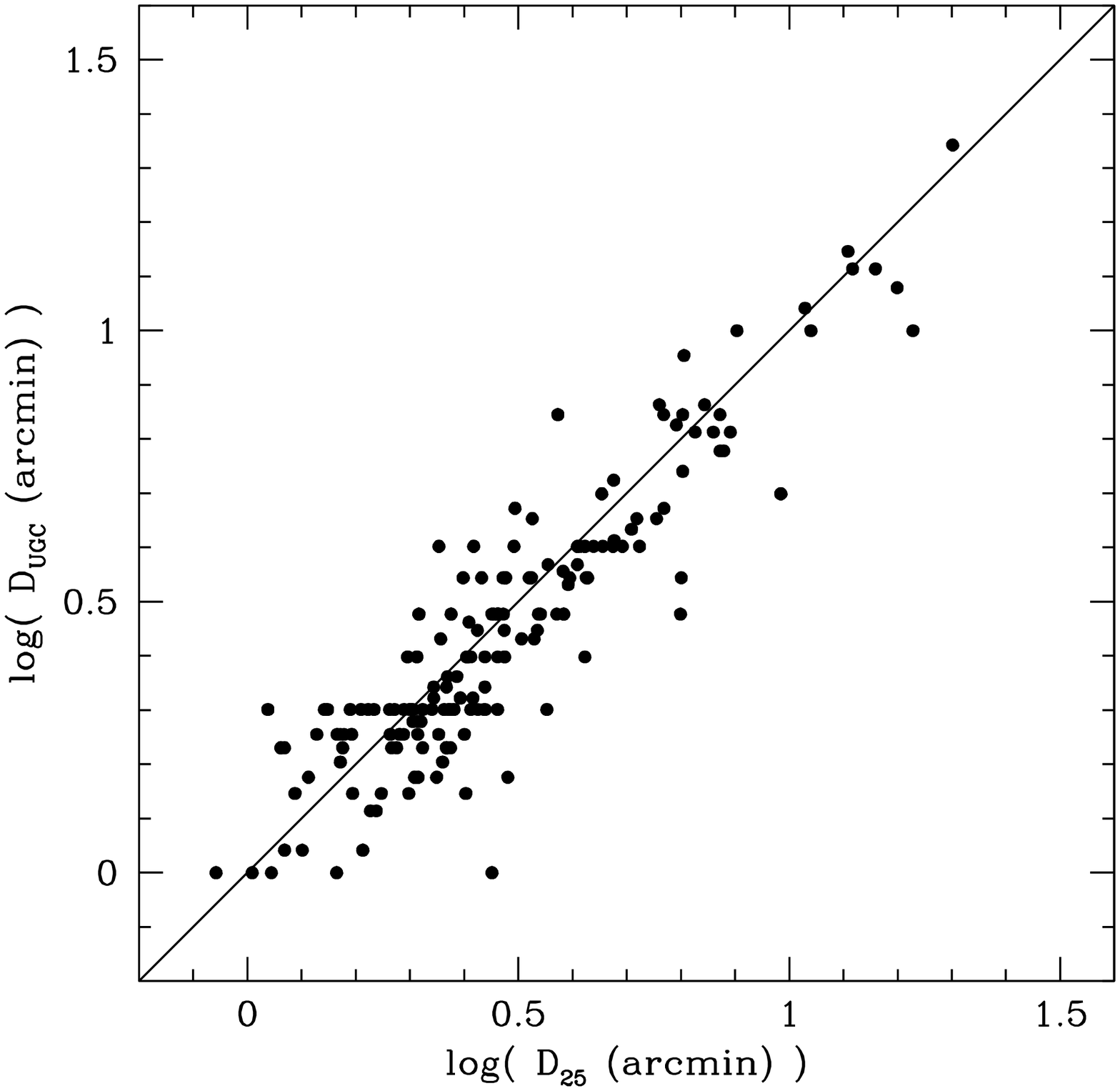}}
\caption{Comparison of $D_{25}$ with the optical diameter as listed in
  the UGC for the $R$-band. The solid line is the line of equality.}
\label{diamcomp}
\vspace{1cm}
\resizebox{\hsize}{!}{\includegraphics{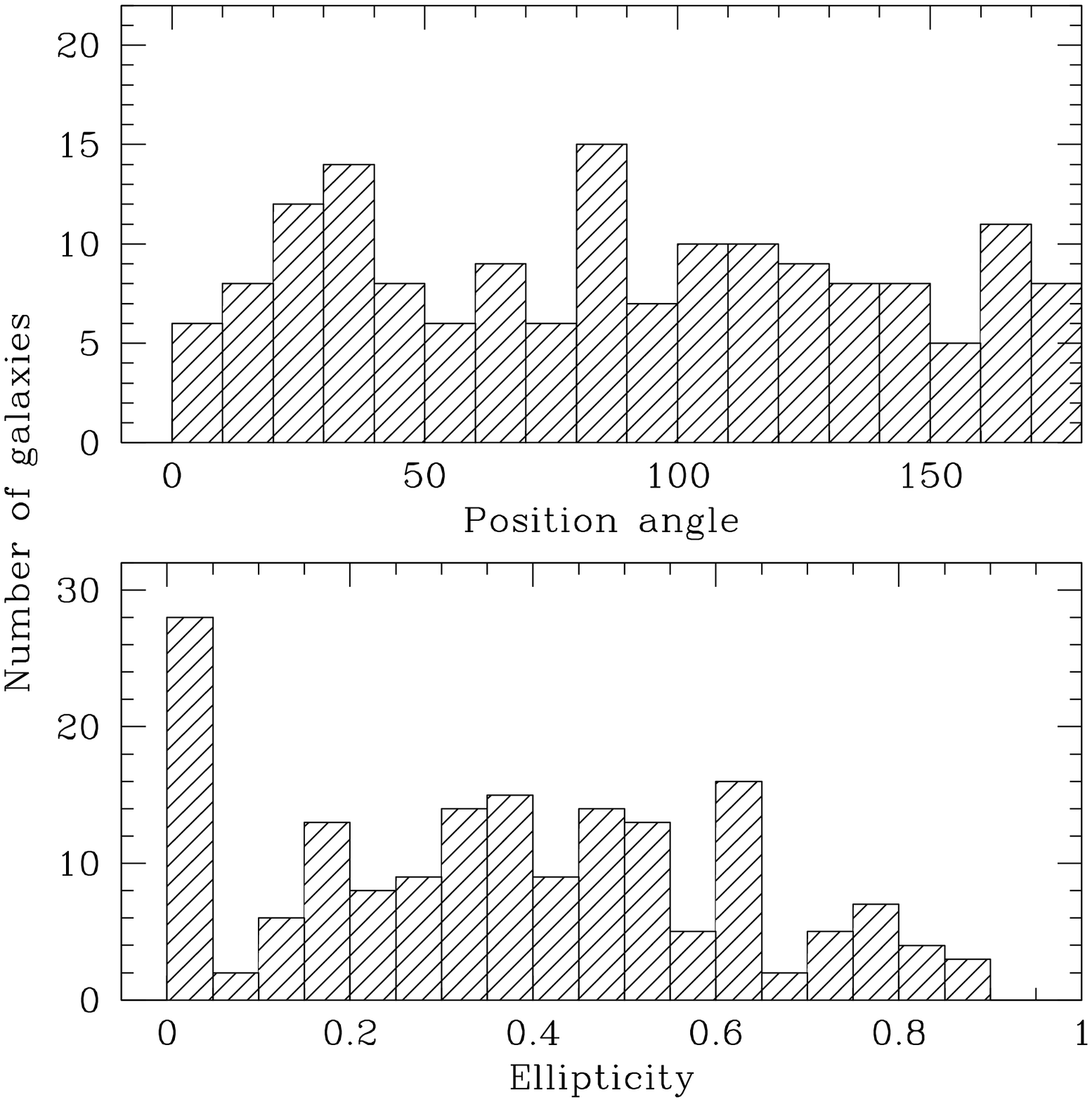}}
\vspace{-0.6cm}
\caption{The distribution of position angle (top panel) and of
  ellipticities (bottom panel) of the galaxies in our sample.}
\label{histpaell}
\vspace{0.2cm}
\end{figure}

Fig.~\ref{ellcomp} shows the comparison of our ellipticities with
those from the UGC. The UGC ellipticities have been determined from
the major and minor-axis diameters listed in the UGC. There is good
general agreement, but with a large scatter, as was to be expected
because our ellipticities have been determined at fainter surface
brightness levels, and because the UGC ellipticities are based on eye
estimates.

\begin{figure}[ht]
\resizebox{\hsize}{!}{\includegraphics{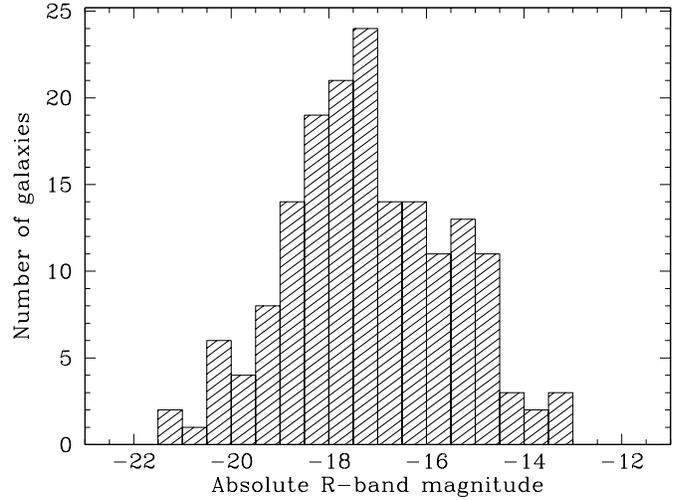}}
\caption{Histogram of absolute $R$-band magnitudes for the galaxies in
  our sample.}
\label{histmabs}
\vspace{1cm}
\end{figure}

In Fig.~\ref{magcompb} the $m_{25}^B$ is compared with the $B$-band
magnitude given in the UGC catalog. Generally, the agreement is good
for $m_{25}^B\la 14.0$. For fainter galaxies, the UGC magnitudes are
systematically too faint. This effect was already noted by Thuan \&
Seitzer (1979). The difference becomes even more pronounced if the UGC
magnitudes are compared with the total magnitude $m_\mathrm{tot}$, where
the difference may be as high as two magnitudes.

A comparison of the UGC $R$-band diameters with the $D_{25}$ diameters
as determined from our $R$-band radial surface brightness profiles
shows good agreement, as can be seen in Fig.~\ref{diamcomp}. This
indicates that the limiting surface brightness at which the UGC
diameter is measured is about 25 $R$-band mag arcsec$^{-2}$. A similar
comparison of the $B$-band diameters shows that the limiting surface
brightness there is about 26.5 $B$-band mag arcsec$^{-2}$, in close
agreement with the value of 26.53 found by Fouqu\'e \& Paturel (1985).

\section{Discussion}
\label{discussion}

As described in Sect.~\ref{sample}, this sample contains a range in
galaxy types and properties. Here we briefly characterize the sample
by showing the distribution of several parameters.

The distribution of ellipticities $\epsilon$ and position angles PA
are given in Fig.~\ref{histpaell}. The ellipticities have been derived
as described in Sect.~\ref{profiles}. The peak at $\epsilon=0$ is
artificial and occurs because the ellipticities were set to zero for
galaxies in which the ellipticities were equal to zero within the
uncertainties.  The true values of the ellipticities for these galaxies
probably spread over a range in ellipticities, up to $\epsilon\sim
0.3$, filling in the depression near $\epsilon=0.1$. This makes the
distribution of ellipticities more or less uniform for small
$\epsilon$. Ellipticities near $\epsilon=1$ are missing because of the
finite thickness of the galaxies.  The distribution of position angles
is consistent with being flat, as is expected if the galaxy
orientations are random.

\begin{figure}[t]
\resizebox{\hsize}{!}{\includegraphics{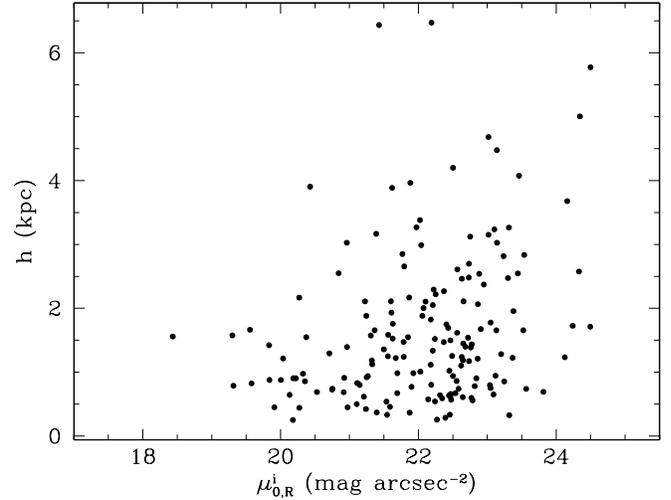}}
\caption{The scale lengths plotted against the inclination corrected
  central disk surface brightnesses for the galaxies in our sample.}
\label{muvshkpc}
\end{figure}

Fig.~\ref{histmabs} shows the distribution over absolute $R$-band
magnitude for the galaxies in our sample. The absolute magnitudes have
been calculated from the total magnitude $m_\mathrm{tot}$ and the
distances listed in Table~\ref{sampletable}, and range from $M_R=-22$
to $-12$.  Most of the galaxies in our sample have low luminosities,
as expected for dwarf galaxies, but about 30\% have $M_R<-18$. These
galaxies are not true dwarf galaxies, but form a mixed bag of large
low surface brightness galaxies (similar to the galaxies studied by
e.g., de Blok \etal\ 1995 and Sprayberry \etal\
1995) and bright interacting galaxies. These galaxies were included in
the sample because our selection is largely based on morphological
type.

Fig.~\ref{muvshkpc} shows this point more clearly. In this figure, for
each galaxy the inclination corrected central disk surface brightness
$\mu_0^i$ is plotted against the scale length. About 30\% of the
galaxies have scale lengths larger than 2 kpc. These are the galaxies
that are more typical of the large low surface brightness galaxies. De
Blok \etal\ (1995) find that these galaxies typically have scale
lengths between 2 and 6 kpc, and surface brightnesses fainter than
$\mu_R=22$ mag arcsec$^{-2}$. Note that the selection on late type has
not included galaxies with both large scale length and high central
surface brightness.

\begin{figure}[t]
\resizebox{\hsize}{!}{\includegraphics{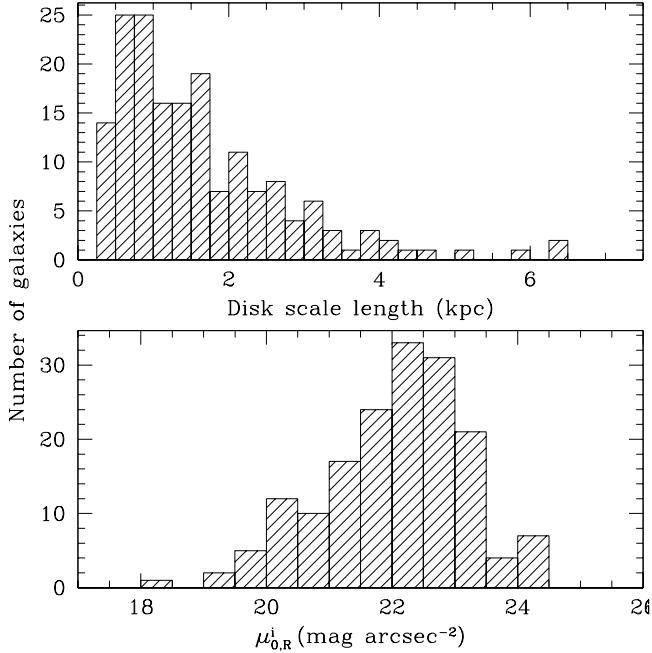}}
\caption{The distribution of scale lengths (top panel) and of
  inclination corrected central surface brightnesses for the galaxies
  in our sample.}
\label{histmuh}
\end{figure}

Finally, Fig.~\ref{histmuh} shows the distribution of scale lengths
and of inclination corrected central disk surface brightnesses. The
galaxies in our sample generally have small scale lengths. The range
in surface brightnesses is large, spanning from  $\mu_{0,R}^i=19$ to 25
mag arcsec$^{-2}$.

\section{Summary}
\label{summary}

We provide $R$-band surface photometry for 171 late-type galaxies.  We
have obtained magnitudes, diameters, central surface brightnesses,
extrapolated central surface brightnesses and scale lengths. For all
galaxies we have presented elliptically-averaged surface brightness
profiles, and profiles showing the run of ellipticity, position angle
profiles and isophote center with radius.  Flatfielding achieved a
median accuracy of 0.22\%, making the surface brightness and
structural profiles reliable to about 26 $R$ mag arcsec$^{-2}$ in most
cases.  The zero point uncertainty is less than 0.1 mag for 77\% of
the sample, and less than 0.2 mag for 89\% of the sample.  The
photometric accuracy is reliably quantified for each galaxy by the
$1\sigma$ errors in the photometric calibration.  Comparison of our
surface brightness profiles and photometric zero points with data from
other authors shows that, for at least 85\% of the sample, we obtain
agreement consistent with the quoted photometric uncertainties.

\acknowledgements

We are grateful to Tjeerd van Albada for valuable comments.  We thank
Rolf Jansen for making available his data before publication.  Travel
grant support for this work was provided by the E.U. ANTARES
Astrophysics Network, under contract CHRX-CT.93.0359, and by the Leids
Kerkhoven-Bosscha Fonds.  This research has made use of the NASA/IPAC
Extragalactic Database (NED) which is operated by the Jet Propulsion
Laboratory, California Institute of Technology, under contract with
the National Aeronautics and Space Administration.  IRAF is
distributed by the National Optical Astronomy Observatories, which are
operated by the Association of Universities for Research in Astronomy,
Inc., under cooperative agreement with the National Science
Foundation. The Digitized Sky Survey (DSS) was produced at the Space
Telescope Science Institute under U.S. Government grant NAG W-2166.

\endacknowledgements

\clearpage


\appendix
{\parskip=0pt
\section{Tables}
\label{thetables}

\subsection[Global properties]{Table~\ref{sampletable} -- Global
  properties}

\noindent {\it Column} (1) gives the UGC number.

\noindent {\it Column} (2) provides other common names, in this order:
NGC, DDO (van den Bergh 1959, 1966), IC, Arp (Arp 1966), CGCG.  At most
two other names are given.

\noindent {\it Columns} (3) and (4) give the equatorial coordinates
(2000) derived from the optical images, as described in
Sect.~\ref{observations}.

\noindent {\it Column} (5) gives the morphological type according to
the RC3, using the same coding.

\noindent {\it Column} (6) gives the absolute $B$-band magnitude,
calculated from the apparent photographic magnitude as given in the
RC3, and the distance as given in column 7.

\noindent {\it Column} (7) provides the adopted distance.  Where
possible stellar distance indicators have been used, mostly Cepheids
and brightest stars.  If these were not available, a distance based on
group membership was used.  If these were not available either, the
distance was calculated from the HI systemic velocity following the
prescription given in Kraan-Korteweg (1986), with an adopted Hubble
constant of $H_0=75$ \kms\ Mpc$^{-1}$.  A full list of published
distances for the galaxies in this sample, updated to the beginning of
1998, is given in Table~\ref{disttable}. For a discussion of the
distance uncertainties, see Sect.~\ref{thedist}.

\noindent {\it Column} (8), (9) and (10) give the heliocentric velocity,
the HI line width at the 50\% level and the HI mass in units of $10^8$
M$_\odot$, respectively, all as given in the RC3.

\noindent {\it Column} (11) gives the Galactic extinction in the $R$-band,
derived from the $A_B$ value according to Burstein \& Heiles (1984)
assuming $A_B/A_R$ of 1.77 (Rieke \& Lebofsky 1985).

\subsection[Adopted distances]{Table~\ref{disttable} -- Adopted distances}

\noindent {\it Column} (1) gives the UGC number

\noindent {\it Column} (2) gives the distance as derived from the HI
systemic velocity, assuming a Hubble constant $H_0=75$ \kms Mpc$^{-1}$
and following the prescription given in Kraan-Korteweg
(1986) to correct for the Virgocentric inflow.

\noindent {\it Column} (3) gives the distance(s) found in the
literature.

\noindent {\it Column} (4) gives the reference for each distance.

\noindent {\it Column} (5) gives a brief description of the method
used in the original paper to derive the distance.

\noindent {\it Column} (6) gives the distance we have adopted. The
distance uncertainties are discussed in Sect.~\ref{thedist}.

\subsection[List of observations]{Table~\ref{obslistR} and~\ref{obslistB} --
List of observations}

\noindent {\it Column} (1) lists the UGC number.

\noindent {\it Column} (2) gives the observing date.

\noindent {\it Column} (3) gives the exposure time.

\noindent {\it Column} (4) gives an estimate of the seeing, as described
in Sect.~\ref{observations}.

\noindent {\it Column} (5) gives the photometric accuracy $\sigma_{\rm
phot}$.  This is the residual obtained in the photometric solution, as
described in Sect.~\ref{observations}.

\subsection[Isophotal and photometric parameters]{Table~\ref{globpropR}
and~\ref{globpropB} -- isophotal and photometric parameters}

\noindent {\it Column} (1) lists the UGC number.

\noindent {\it Column} (2) and (3) give the ellipticity $\epsilon$ and the
position angle P.A.

\noindent {\it Column} (4) gives the apparent magnitude
$m_{25}$ within the 25 mag arcsec$^{-2}$ isophote.

\noindent {\it Column} (5) gives the apparent limiting
magnitude $m_{\rm lim}$.  This is the apparent magnitude within the
limiting surface brightness $\mu_{\rm lim}$ given in column (7).

\noindent {\it Column} (6) gives the extrapolated apparent
magnitude $m_{\rm ext}$. The extrapolation is described in
Sect.~\ref{parameters}.

\noindent {\it Column} (7) gives the limiting surface brightness
$\mu_{\rm lim}$, which corresponds to $3\sigma$ above sky.  $\sigma$
is our estimate of the $1\sigma$ error in the sky, as described see
Sect.~\ref{observations}.

\noindent {\it Column} (8) gives the central surface brightness $\mu_c$,
as determined from the luminosity profile (see
Sect.~\ref{profiles}).  $\mu_c$ has been corrected for Galactic
foreground extinction, but not for inclination.

\noindent {\it Column} (9) gives the extrapolated disk central surface
brightness $\mu_0$ as obtained from an exponential fit to the surface
brightness profile. $\mu_0$ has been corrected for Galactic foreground
extinction, but not for inclination.

\noindent {\it Column} (10) gives the absolute magnitude M$_R$ or
M$_B$.  The absolute magnitude has been calculated from $m_{\rm ext}$
and the distance as listed in \ref{disttable}. These values are
corrected for Galactic foreground extinction.

\noindent {\it Column} (11) gives the uncertainty in the photometry,
$\sigma_{\rm phot}$, which is the residual obtained in the photometric
solution, as described in Sect.~\ref{observations}. For galaxies
with $\sigma_{\rm phot}>0.25$ mag the photometric parameters are only
given to the nearest tenth of a magnitude.

\noindent {\it Column} (12) gives the scale length $h$ in arcsec, measured
along
the major axis, as determined from an exponential fit to surface
brightness profile.

\noindent {\it Column} (13) and (14) give the isophotal diameters,
$d_{25}$ and $d_{26.5}$ at the 25 and 26.5 mag arcsec$^{-2}$ isophote.
These diameters have been corrected for Galactic foreground extinction
and are measured along the major axis. The diameters are given in
units of arcsec.

\noindent {\it Column} (15), (16) and (17) give the radii $r_{20}$,
$r_{50}$ and $r_{80}$ within which 20\%, 50\% and 80\% of the light is
contained. These radii have been measured along the major axis, and
are given in units of arcsec.

\bigskip\centerline{--- {\it Notes} ---}\medskip

\noindent $^a$~~These galaxies are at very low Galactic latitude and
therefore
Burstein \& Heiles (1984) do not give Galactic foreground extinctions.
The galaxies concerned are: UGC~192, UGC~3272 and UGC~3390. For these
galaxies the magnitudes and surface brightnesses are not
corrected for Galactic extinction.

\noindent $^b$~~These galaxies have been observed during nights with
occasional thin cirrus clouds. Therefore, in some cases the
$1\sigma$ photometric error may be an underestimate of the true photometric
error, as was found in Sect.~\ref{comparison}.

}\clearpage

{\renewcommand{\baselinestretch}{0.94}

\setlength{\tabcolsep}{3pt}
\def\ruleit#1{\vrule width#1 height0pt depth0pt}
{
\begin{table*}
\null\vspace{-1pt}
\caption[]{Global properties\\}
\label{sampletable}
\begin{flushleft}
\begin{tabular}{rlrrrrrrcrrrrrr}
\hline\noalign{\smallskip}
 UGC & \hfil Other names & \multicolumn{3}{c}{R.A. (2000)}& \multicolumn{3}{c}{Dec. (2000)} & Type & $M_B$ \hfil & D \hfil & $\varv_{\rm HI}$ \hfil & $W_{50}$ \hfil & M$_{\rm HI}$ \hfil & $A_R$ \hfil \\
  &  & \multicolumn{3}{c}{$^h$ $^m$ $^s$}& \multicolumn{3}{c}{$^\circ$ $'$ $''$} &  & mag \hfil & Mpc \hfil & km/s \hfil & km/s \hfil & $10^8$ M$_\odot$ \hfil & mag \hfil \\
 (1) & \hfil (2) & \multicolumn{3}{c}{(3)}& \multicolumn{3}{c}{(4)} & (5) & (6) \hfil & (7) \hfil & (8) \hfil & (9) \hfil & (10) \hfil & (11) \hfil \\
\noalign{\smallskip}\hline\noalign{\smallskip}
\omit\ruleit{1.0cm}&\ruleit{3.4cm}&\ruleit{0.4cm}&\ruleit{0.4cm}&\ruleit{0.6cm}&\ruleit{0.6cm}&\ruleit{0.4cm}&\ruleit{0.4cm}&\ruleit{1.4cm}&\ruleit{0.9cm}&\ruleit{0.8cm}&\ruleit{0.9cm}&\ruleit{0.7cm}&\ruleit{0.7cm}&\ruleit{0.9cm}\\
192 & IC 10 & 0 & 20 & 20.0 & 59 & 17 & 56 & .IB.9\$. & -- & 1.0 & -343 & 63 & 1.33 & -- \\
731 & DDO 9 & 1 & 10 & 43.6 & 49 & 36 & 4 & .I..9*. & -- & 8.0 & 639 & 130 & 5.99 & 0.41 \\
1249 & IC 1727 & 1 & 47 & 29.9 & 27 & 19 & 56 & .SBS9.. & -17.59 & 7.5 & 338 & 121 & 8.2 & 0.14 \\
1281 &  & 1 & 49 & 31.9 & 32 & 35 & 24 & .S..8.. & -15.82 & 5.5 & 157 & 116 & 2.83 & 0.08 \\
1438 & NGC 746 & 1 & 57 & 51.1 & 44 & 55 & 6 & .I..9.. & -17.11 & 13.3 & 712 & 105 & 8.38 & 0.31 \\
1501 & NGC 784 & 2 & 01 & 16.9 & 28 & 50 & 10 & .SB.8*/ & -16.86 & 5.7 & 198 & 96 & 5.05 & 0.12 \\
1547 & DDO 17 & 2 & 03 & 20.4 & 22 & 02 & 31 & .IB.9.. & -17.22 & 20.2 & 2644 & 138 & 25.9 & 0.15 \\
1551 &  & 2 & 03 & 37.6 & 24 & 04 & 32 & .SB?... & -18.02 & 20.2 & 2671 & 118 & 14.1 & 0.19 \\
1865 & DDO 19 & 2 & 24 & 59.9 & 36 & 02 & 15 & .S..9*. & -- & 10.1 & 580 & 76 & 3.22 & 0.12 \\
2014 & DDO 22 & 2 & 32 & 54.0 & 38 & 40 & 35 & .I..9*. & -- & 10.1 & 565 & 49 & 1.94 & 0.1 \\
2017 &  & 2 & 32 & 45.4 & 28 & 50 & 29 & .I..9.. & -- & 16.2 & 1014 & 84 & 9.09 & 0.16 \\
2023 & DDO 25 & 2 & 33 & 18.2 & 33 & 29 & 28 & .I..9*. & -15.61 & 10.1 & 606 & 39 & 4.06 & 0.19 \\
2034 & DDO 24 & 2 & 33 & 43.1 & 40 & 31 & 43 & .I..9.. & -15.71 & 10.1 & 579 & 45 & 8.17 & 0.1 \\
2053 & DDO 26 & 2 & 34 & 29.2 & 29 & 45 & 4 & .I..9.. & -15.35 & 11.8 & 1029 & 59 & 5.59 & 0.2 \\
2455 & NGC 1156 & 2 & 59 & 42.2 & 25 & 14 & 15 & .IBS9.. & -17.45 & 7.8 & 382 & 67 & 10.7 & 0.37 \\
2603 &  & 3 & 19 & 14.6 & 81 & 20 & 47 & .I..9.. & -- & 38.5 & 2519 & 121 & 43.1 & 0.39 \\
2800 &  & 3 & 40 & 03.8 & 71 & 24 & 20 & .I..9?. & -- & 20.6 & 1176 & 208 & 29.3 & 0.77 \\
3056 & NGC 1569, Arp 210 & 4 & 30 & 49.6 & 64 & 50 & 53 & .IB.9.. & -14.8 & 2.2 & -88 & 74 & 1.12 & 1.15 \\
3060 & NGC 1560 & 4 & 32 & 48.3 & 71 & 52 & 50 & .SAS7./ & -16.15 & 3.9 & -35 & 125 & 10.3 & 0.35 \\
3137 &  & 4 & 46 & 15.0 & 76 & 25 & 6 & .S?.... & -16.22 & 18.4 & 993 & 222 & 35.4 & 0.31 \\
3144 & DDO 33 & 4 & 47 & 54.5 & 74 & 55 & 46 & .IB.9.. & -16.24 & 19.5 & 1636 & 142 & 22.8 & 0.32 \\
3273 &  & 5 & 17 & 44.9 & 53 & 33 & 1 & .S..9.. & -15.62 & 12.2 & 616 & 185 & 19 & -- \\
3317 & DDO 38 & 5 & 33 & 37.7 & 73 & 43 & 30 & .I..9.. & -- & 19.5 & 1239 & 107 & 14.3 & 0.26 \\
3371 & DDO 39 & 5 & 56 & 37.5 & 75 & 19 & 0 & .I..9*. & -- & 12.8 & 816 & 129 & 12.6 & 0.27 \\
3384 &  & 6 & 01 & 37.3 & 73 & 07 & 1 & .S..9*. & -- & 19.5 & 1086 & 78 & 17.5 & 0.3 \\
3390 &  & 6 & 02 & 05.3 & 36 & 06 & 17 & .SX.8.. & -- & 23.2 & 1517 & 202 & 27.2 & -- \\
3475 &  & 6 & 30 & 28.9 & 39 & 30 & 15 & .S..9*. & -15.33 & 9.3 & 486 & 167 & 4.88 & 0.66 \\
3647 & DDO 40 & 7 & 04 & 50.2 & 56 & 31 & 9 & .IB.9.. & -- & 18.4 & 1384 & 50 & 10.2 & 0.08 \\
3698 &  & 7 & 09 & 18.8 & 44 & 22 & 49 & .I..9*. & -- & 8.5 & 426 & 50 & 1.76 & 0.23 \\
3711 & NGC 2337 & 7 & 10 & 13.6 & 44 & 27 & 27 & .IB.9.. & -16.71 & 8.6 & 434 & 144 & 6.86 & 0.21 \\
3817 &  & 7 & 22 & 44.6 & 45 & 06 & 31 & .I..9*. & -- & 8.7 & 438 & 43 & 1.88 & 0.19 \\
3851 & NGC 2366, DDO 42 & 7 & 28 & 52.9 & 69 & 12 & 45 & .IBS9.. & -15.98 & 3.4 & 100 & 96 & 7.92 & 0.1 \\
3860 & DDO 43 & 7 & 28 & 17.5 & 40 & 46 & 11 & .I..9.. & -14.05 & 6.8 & 354 & 38 & 1.3 & 0.12 \\
3966 & DDO 46 & 7 & 41 & 26.1 & 40 & 06 & 39 & .I..9.. & -- & 6.0 & 361 & 71 & 1.69 & 0.12 \\
4173 &  & 8 & 07 & 10.5 & 80 & 07 & 38 & .I..9*. & -15.82 & 16.8 & 862 & 0 & 20.2 & 0.06 \\
4274 & NGC 2537, Arp 6 & 8 & 13 & 14.7 & 45 & 59 & 24 & .SBS9P. & -16.98 & 6.6 & 447 & 96 & 2.06 & 0.08 \\
4278 & IC 2233 & 8 & 13 & 58.8 & 45 & 44 & 36 & .SBS7*/ & -16.62 & 10.5 & 563 & 180 & 13.3 & 0.08 \\
4305 & Arp 268 & 8 & 19 & 07.4 & 70 & 43 & 24 & .I..9.. & -16.75 & 3.4 & 158 & 66 & 8.21 & 0.05 \\
4325 & NGC 2552 & 8 & 19 & 19.7 & 50 & 00 & 32 & .SAS9\$. & -17.42 & 10.1 & 519 & 134 & 7.52 & 0.1 \\
4426 & DDO 52 & 8 & 28 & 28.3 & 41 & 51 & 22 & .I..9*. & -- & 6.4 & 393 & 83 & 1 & 0.07 \\
4483 &  & 8 & 37 & 03.5 & 69 & 46 & 32 & .I..9*. & -12.77 & 3.6 & 156 & 49 & 0.41 & 0.08 \\
4499 &  & 8 & 37 & 41.4 & 51 & 39 & 9 & .SX.8.. & -17.06 & 13.0 & 692 & 126 & 12 & 0.07 \\
4543 &  & 8 & 43 & 21.7 & 45 & 44 & 10 & .SA.8.. & -18.1 & 30.3 & 1960 & 108 & 60 & 0.06 \\
4660 &  & 8 & 54 & 24.2 & 34 & 33 & 22 & .S..9*. & -- & 32.9 & 2203 & 61 & 18.6 & 0.03 \\
4704 &  & 8 & 59 & 00.3 & 39 & 12 & 39 & .S..8*. & -15.03 & 10.2 & 596 & 129 & 5.56 & 0.03 \\
4945 &  & 9 & 22 & 25.8 & 75 & 46 & 2 & .I..9.. & -- & 6.7 & 659 & 34 & 0.38 & 0 \\
5040 &  & 9 & 27 & 36.4 & 28 & 47 & 58 & .I..9?. & -19.81 & 58.0 & 4149 & 63 & 70.2 & 0 \\
5139 & DDO 63 & 9 & 40 & 28.8 & 71 & 10 & 57 & .IXS9.. & -14.65 & 6.8 & 136 & 26 & 4.88 & 0.03 \\
5221 & NGC 2976 & 9 & 47 & 14.9 & 67 & 55 & 2 & .SA.5P. & -17.3 & 4.5 & 3 & 97 & 2.57 & 0.06 \\
5272 & DDO 64 & 9 & 50 & 22.7 & 31 & 29 & 14 & .I..9.. & -14.82 & 6.1 & 520 & 82 & 1.52 & 0.02 \\
5322 & NGC 3034 & 9 & 55 & 52.7 & 69 & 40 & 52 & .I.0../ & -19.61 & 5.9 & 203 & 146 & 18.2 & 0.07 \\
5336 & DDO 66 & 9 & 57 & 31.7 & 69 & 02 & 48 & .I..9.. & -- & 3.6 & 46 & 69 & 3.28 & 0.08 \\
5340 & DDO 68 & 9 & 56 & 45.7 & 28 & 49 & 34 & .I..9P* & -14.82 & 6.1 & 503 & 78 & 2.6 & 0.02 \\
5364 & DDO 69 & 9 & 59 & 25.2 & 30 & 44 & 48 & .IB.9.. & -13.9 & 2.2 & 20 & 33 & 0.74 & 0.04 \\
5398 & NGC 3077 & 10 & 03 & 19.1 & 68 & 44 & 5 & .I.0.P. & -16.98 & 3.6 & 14 & 65 & 10.5 & 0.12 \\
5414 & NGC 3104, Arp 264 & 10 & 03 & 58.4 & 40 & 45 & 24 & .IXS9.. & -16.39 & 10.0 & 612 & 100 & 5.75 & 0.02 \\
5455 &  & 10 & 08 & 50.3 & 70 & 38 & 3 & .I..9.. & -- & 22.6 & 1290 & 57 & 9.55 & 0.1 \\
\noalign{\smallskip}\hline
\end{tabular}
\end{flushleft}
\vfill
\end{table*}
\addtocounter{table}{-1}
\begin{table*}
\null\vspace{-1pt}
\caption[]{-- Continued}
\begin{flushleft}
\begin{tabular}{rlrrrrrrcrrrrrr}
\hline\noalign{\smallskip}
 UGC & \hfil Other names & \multicolumn{3}{c}{R.A. (2000)}& \multicolumn{3}{c}{Dec. (2000)} & Type & $M_B$ \hfil & D \hfil & $\varv_{\rm HI}$ \hfil & $W_{50}$ \hfil & M$_{\rm HI}$ \hfil & $A_R$ \hfil \\
  &  & \multicolumn{3}{c}{$^h$ $^m$ $^s$}& \multicolumn{3}{c}{$^\circ$ $'$ $''$} &  & mag \hfil & Mpc \hfil & km/s \hfil & km/s \hfil & $10^8$ M$_\odot$ \hfil & mag \hfil \\
 (1) & \hfil (2) & \multicolumn{3}{c}{(3)}& \multicolumn{3}{c}{(4)} & (5) & (6) \hfil & (7) \hfil & (8) \hfil & (9) \hfil & (10) \hfil & (11) \hfil \\
\noalign{\smallskip}\hline\noalign{\smallskip}
\omit\ruleit{1.0cm}&\ruleit{3.4cm}&\ruleit{0.4cm}&\ruleit{0.4cm}&\ruleit{0.6cm}&\ruleit{0.6cm}&\ruleit{0.4cm}&\ruleit{0.4cm}&\ruleit{1.4cm}&\ruleit{0.9cm}&\ruleit{0.8cm}&\ruleit{0.9cm}&\ruleit{0.7cm}&\ruleit{0.7cm}&\ruleit{0.9cm}\\
5478 & DDO 73 & 10 & 09 & 31.5 & 30 & 09 & 2 & .I..9.. & -17.03 & 18.5 & 1378 & 55 & 5.83 & 0.03 \\
5612 & DDO 77 & 10 & 24 & 07.2 & 70 & 52 & 55 & .SBS8.. & -16.56 & 13.6 & 1011 & 146 & 10 & 0.11 \\
5666 & IC 2574 & 10 & 28 & 23.0 & 68 & 24 & 56 & .SXS9.. & -17.03 & 3.7 & 47 & 115 & 13.3 & 0.03 \\
5688 & DDO 80 & 10 & 30 & 25.0 & 70 & 03 & 3 & .SB.9*. & -- & 30.2 & 1921 & 54 & 32.7 & 0 \\
5706 &  & 10 & 31 & 10.8 & 34 & 30 & 13 & .I..9.. & -- & 24.5 & 1494 & 36 & 5.78 & 0.01 \\
5721 & NGC 3274 & 10 & 32 & 17.3 & 27 & 40 & 8 & .SX.7?. & -15.95 & 6.7 & 537 & 157 & 6.85 & 0.03 \\
5740 &  & 10 & 34 & 47.0 & 50 & 46 & 11 & .SX.9.. & -15.27 & 11.9 & 651 & 117 & 4.52 & 0 \\
5764 & DDO 83 & 10 & 36 & 43.5 & 31 & 32 & 50 & .IBS9*. & -15.36 & 9.0 & 586 & 101 & 2.56 & 0.02 \\
5829 & DDO 84 & 10 & 42 & 43.2 & 34 & 27 & 1 & .I..9.. & -15.76 & 9.0 & 630 & 75 & 12.7 & 0.02 \\
5846 & DDO 86 & 10 & 44 & 29.9 & 60 & 22 & 6 & .I..9.. & -15.69 & 13.2 & 1019 & 44 & 7.06 & 0 \\
5860 & NGC 3353 & 10 & 45 & 22.4 & 55 & 57 & 37 & .S..3\$P & -17.93 & 17.4 & 944 & 96 & 8.25 & 0 \\
5918 & VII Zw 347 & 10 & 49 & 37.0 & 65 & 31 & 48 & .I..9*. & -- & 7.7 & 338 & 62 & 2.45 & 0 \\
5935 & NGC 3396, Arp 270 & 10 & 49 & 55.4 & 32 & 59 & 27 & .IB.9P. & -19.49 & 26.4 & 1625 & 160 & 56.3 & 0 \\
5986 & NGC 3432, Arp 206 & 10 & 52 & 31.3 & 36 & 37 & 9 & .SBS9./ & -17.92 & 8.7 & 616 & 232 & 22.8 & 0 \\
6016 &  & 10 & 54 & 13.0 & 54 & 17 & 14 & .I..9.. & -- & 25.2 & 1493 & 136 & 25 & 0 \\
6021 & NGC 3445 & 10 & 54 & 35.6 & 56 & 59 & 27 & .SXS9.. & -19.66 & 32.1 & 2023 & 102 & 43.3 & 0 \\
6024 & NGC 3448 & 10 & 54 & 38.9 & 54 & 18 & 21 & .I.0... & -19.78 & 23.2 & 1350 & 239 & 79.9 & 0 \\
6126 & NGC 3510 & 11 & 03 & 43.4 & 28 & 53 & 11 & .SBS9./ & -16.75 & 8.9 & 705 & 186 & 9.17 & 0 \\
6151 & DDO 91 & 11 & 05 & 56.4 & 19 & 49 & 35 & .S..9*. & -16.87 & 17.2 & 1331 & 26 & 4.39 & 0 \\
6161 &  & 11 & 06 & 49.2 & 43 & 43 & 23 & .SB.8.. & -16.54 & 12.9 & 758 & 108 & 9.57 & 0 \\
6251 & DDO 92 & 11 & 13 & 26.8 & 53 & 35 & 44 & .SX.9*. & -15.59 & 13.2 & 927 & 44 & 4.79 & 0 \\
6406 & NGC 3657 & 11 & 23 & 55.6 & 52 & 55 & 16 & .SXT5P. & -18.53 & 21.5 & 1215 & 196 & 39.1 & 0 \\
6446 &  & 11 & 26 & 40.4 & 53 & 44 & 50 & .SA.7.. & -16.59 & 12.0 & 645 & 135 & 14.1 & 0 \\
6456 & VII Zw 403 & 11 & 27 & 58.7 & 78 & 59 & 38 & .P..... & -12.88 & 3.0 & -92 & 49 & 0.33 & 0.05 \\
6565 & NGC 3738, Arp 234 & 11 & 35 & 48.6 & 54 & 31 & 28 & .I..9.. & -15.71 & 3.2 & 229 & 78 & 0.73 & 0 \\
6628 &  & 11 & 40 & 05.8 & 45 & 56 & 32 & .SA.9.. & -17.01 & 15.3 & 850 & 40 & 13.3 & 0.01 \\
6682 & DDO 96 & 11 & 43 & 09.0 & 59 & 06 & 24 & .S..9*. & -- & 16.1 & 1326 & 73 & 7.33 & 0 \\
6817 & DDO 99 & 11 & 50 & 54.1 & 38 & 52 & 54 & .I..9.. & -13.71 & 4.02 & 245 & 37 & 1.49 & 0 \\
6840 & DDO 100 & 11 & 52 & 07.0 & 52 & 06 & 29 & .SBT9.. & -15.77 & 15.7 & 1046 & 143 & 8.86 & 0.01 \\
6912 & VII Zw 430 & 11 & 56 & 13.9 & 58 & 11 & 60 & .S?.... & -17.26 & 23.6 & 1357 & 96 & 22.1 & 0 \\
6917 &  & 11 & 56 & 28.7 & 50 & 25 & 42 & .SB.9.. & -17.63 & 16.9 & 910 & 183 & 19.3 & 0.02 \\
6944 & NGC 3995, Arp 313 & 11 & 57 & 44.1 & 32 & 17 & 38 & .SA.9P. & -20.52 & 47.4 & 3254 & 133 & 148 & 0.01 \\
6955 & DDO 105 & 11 & 58 & 29.1 & 38 & 04 & 33 & .IBS9*. & -16.11 & 11.6 & 909 & 144 & 10.5 & 0 \\
6956 & DDO 102 & 11 & 58 & 26.1 & 50 & 55 & 6 & .SBS9.. & -- & 15.7 & 917 & 52 & 6.3 & 0.02 \\
6995 & NGC 4032 & 12 & 00 & 32.9 & 20 & 04 & 25 & .I..9*. & -19.08 & 22.6 & 1269 & 103 & 22.4 & 0.03 \\
7047 & NGC 4068 & 12 & 04 & 02.2 & 52 & 35 & 25 & .IA.9.. & -14.71 & 3.5 & 211 & 51 & 1.04 & 0 \\
7125 &  & 12 & 08 & 42.1 & 36 & 48 & 10 & .S..9.. & -17.24 & 19.5 & 1071 & 132 & 46.9 & 0.01 \\
7151 & NGC 4144 & 12 & 09 & 58.2 & 46 & 27 & 28 & .SXS6\$/ & -15.58 & 3.5 & 267 & 150 & 1.58 & 0 \\
7199 & NGC 4163 & 12 & 12 & 09.0 & 36 & 10 & 6 & .IA.9.. & -14.25 & 3.5 & 164 & 30 & 0.24 & 0 \\
7204 & NGC 4173 & 12 & 12 & 21.2 & 29 & 12 & 29 & .SB.7*. & -18.33 & 20.3 & 1104 & 117 & 40 & 0.01 \\
7232 & NGC 4190 & 12 & 13 & 44.6 & 36 & 38 & 9 & .I..9P. & -14.47 & 3.5 & 230 & 46 & 0.59 & 0 \\
7261 & NGC 4204 & 12 & 15 & 14.2 & 20 & 39 & 34 & .SBS8.. & -16.89 & 9.1 & 861 & 85 & 6.63 & 0.04 \\
7278 & NGC 4214 & 12 & 15 & 39.2 & 36 & 19 & 37 & .IXS9.. & -17.53 & 3.5 & 291 & 62 & 8.09 & 0 \\
7306 & NGC 4236 & 12 & 16 & 42.8 & 69 & 27 & 46 & .SBS8.. & -17.05 & 3.0 & 0 & 162 & 13 & 0.03 \\
7323 & NGC 4242 & 12 & 17 & 30.1 & 45 & 37 & 10 & .SXS8.. & -18.02 & 8.1 & 517 & 115 & 7.06 & 0 \\
7356 &  & 12 & 19 & 09.1 & 47 & 05 & 25 & .I..9?. & -- & 3.5 & 272 & 86 & 1.15 & 0 \\
7399 & NGC 4288, DDO 119 & 12 & 20 & 38.1 & 46 & 17 & 31 & .SBS8.. & -16.25 & 8.4 & 535 & 173 & 6.37 & 0 \\
7408 & DDO 120 & 12 & 21 & 15.4 & 45 & 48 & 47 & .IA.9.. & -15.31 & 8.4 & 462 & 26 & 1.63 & 0 \\
7490 & DDO 122 & 12 & 24 & 25.0 & 70 & 20 & 2 & .SA.9.. & -15.84 & 8.5 & 467 & 69 & 2.85 & 0 \\
7524 & NGC 4395 & 12 & 25 & 48.8 & 33 & 32 & 50 & .SAS9*. & -17.01 & 3.5 & 320 & 109 & 9.54 & 0.01 \\
7534 & DDO 123 & 12 & 26 & 08.1 & 58 & 19 & 20 & .IB.9.. & -15.14 & 8.5 & 723 & 55 & 5.13 & 0 \\
7559 & DDO 126 & 12 & 27 & 05.5 & 37 & 08 & 34 & .IB.9.. & -12.82 & 3.2 & 218 & 59 & 0.61 & 0 \\
7577 & DDO 125 & 12 & 27 & 42.0 & 43 & 29 & 37 & .I..9.. & -13.81 & 3.5 & 196 & 28 & 0.71 & 0 \\
7592 & NGC 4449 & 12 & 28 & 11.3 & 44 & 05 & 36 & .IB.9.. & -17.61 & 3.5 & 201 & 136 & 12.4 & 0 \\
7599 & DDO 127 & 12 & 28 & 27.9 & 37 & 14 & 2 & .S..9.. & -12.71 & 3.5 & 278 & 66 & 0.31 & 0 \\
7603 & NGC 4455 & 12 & 28 & 44.1 & 22 & 49 & 18 & .SBS7?/ & -16.06 & 6.8 & 644 & 132 & 3.64 & 0.03 \\
7608 & DDO 129 & 12 & 28 & 44.8 & 43 & 13 & 33 & .I..9.. & -- & 8.4 & 538 & 60 & 4.88 & 0 \\
\noalign{\smallskip}\hline
\end{tabular}
\end{flushleft}
\vfill
\end{table*}
\addtocounter{table}{-1}
\begin{table*}
\null\vspace{-1pt}
\caption[]{-- Continued}
\begin{flushleft}
\begin{tabular}{rlrrrrrrcrrrrrr}
\hline\noalign{\smallskip}
 UGC & \hfil Other names & \multicolumn{3}{c}{R.A. (2000)}& \multicolumn{3}{c}{Dec. (2000)} & Type & $M_B$ \hfil & D \hfil & $\varv_{\rm HI}$ \hfil & $W_{50}$ \hfil & M$_{\rm HI}$ \hfil & $A_R$ \hfil \\
  &  & \multicolumn{3}{c}{$^h$ $^m$ $^s$}& \multicolumn{3}{c}{$^\circ$ $'$ $''$} &  & mag \hfil & Mpc \hfil & km/s \hfil & km/s \hfil & $10^8$ M$_\odot$ \hfil & mag \hfil \\
 (1) & \hfil (2) & \multicolumn{3}{c}{(3)}& \multicolumn{3}{c}{(4)} & (5) & (6) \hfil & (7) \hfil & (8) \hfil & (9) \hfil & (10) \hfil & (11) \hfil \\
\noalign{\smallskip}\hline\noalign{\smallskip}
\omit\ruleit{1.0cm}&\ruleit{3.4cm}&\ruleit{0.4cm}&\ruleit{0.4cm}&\ruleit{0.6cm}&\ruleit{0.6cm}&\ruleit{0.4cm}&\ruleit{0.4cm}&\ruleit{1.4cm}&\ruleit{0.9cm}&\ruleit{0.8cm}&\ruleit{0.9cm}&\ruleit{0.7cm}&\ruleit{0.7cm}&\ruleit{0.9cm}\\
7648 & NGC 4485, Arp 269 & 12 & 30 & 31.4 & 41 & 42 & 0 & .IBS9P. & -16.74 & 7.1 & 493 & 139 & 3.48 & 0 \\
7651 & NGC 4490 & 12 & 30 & 36.3 & 41 & 38 & 36 & .IBS9P. & -19.29 & 8.4 & 578 & 173 & 52.9 & 0 \\
7673 & DDO 131 & 12 & 31 & 58.7 & 29 & 42 & 34 & .I..9.. & -- & 8.4 & 642 & 55 & 1.56 & 0.02 \\
7690 &  & 12 & 32 & 26.9 & 42 & 42 & 17 & .I..9*. & -15.99 & 7.9 & 537 & 88 & 3.07 & 0 \\
7698 & DDO 133 & 12 & 32 & 54.5 & 31 & 32 & 21 & .I..9.. & -14.71 & 3.5 & 333 & 53 & 1.2 & 0.02 \\
7723 & NGC 4534 & 12 & 34 & 05.4 & 35 & 31 & 7 & .SAS8*. & -17.36 & 11.3 & 802 & 123 & 19.4 & 0 \\
7831 & NGC 4605 & 12 & 40 & 00.6 & 61 & 36 & 31 & .SBS5P. & -17.64 & 5.2 & 143 & 133 & 3.69 & 0 \\
7853 & NGC 4618, Arp 23 & 12 & 41 & 32.5 & 41 & 09 & 2 & .SBT9.. & -18.11 & 7.8 & 543 & 97 & 12.3 & 0 \\
7861 & NGC 4625, IC 3675 & 12 & 41 & 52.5 & 41 & 16 & 20 & .SXT9P. & -16.91 & 9.0 & 611 & 66 & 5.86 & 0 \\
7866 & IC 3687 & 12 & 42 & 15.0 & 38 & 30 & 13 & .IXS9.. & -13.9 & 4.8 & 359 & 46 & 1.16 & 0 \\
7872 &  & 12 & 41 & 54.1 & 75 & 18 & 27 & .I..9.. & -- & 30.7 & 1887 & 69 & 18.6 & 0.02 \\
7907 & NGC 4656 & 12 & 43 & 58.3 & 32 & 10 & 18 & .SBS9P. & -18.93 & 7.9 & 640 & 139 & 49.5 & 0 \\
7916 & I Zw 42 & 12 & 44 & 25.0 & 34 & 23 & 13 & .I..9.. & -- & 8.4 & 607 & 57 & 3.6 & 0.01 \\
7949 & DDO 147 & 12 & 47 & 00.1 & 36 & 28 & 41 & .I..9*. & -- & 4.8 & 333 & 30 & 1 & 0.02 \\
7971 & NGC 4707, DDO 150 & 12 & 48 & 22.5 & 51 & 09 & 55 & .S..9*. & -15.11 & 8.4 & 467 & 64 & 2.49 & 0.01 \\
8005 & NGC 4747 & 12 & 51 & 46.0 & 25 & 46 & 35 & .SB6\$/P & -18.91 & 21.8 & 1188 & 127 & 31 & 0.02 \\
8024 & NGC 4789A, DDO 154 & 12 & 54 & 05.4 & 27 & 08 & 56 & .IBS9.. & -14.9 & 4.8 & 376 & 85 & 4.47 & 0.01 \\
8098 & NGC 4861, IC 3961 & 12 & 59 & 02.0 & 34 & 51 & 41 & .SBS9*. & -- & 12.8 & 846 & 86 & 16.6 & 0 \\
8188 & IC 4182 & 13 & 05 & 49.5 & 37 & 36 & 24 & .SAS9.. & -15.35 & 4.7 & 321 & 40 & 3.25 & 0 \\
8201 & VII Zw 499 & 13 & 06 & 25.6 & 67 & 42 & 19 & .I..9.. & -14.84 & 4.9 & 37 & 44 & 1.66 & 0.02 \\
8286 & NGC 5023 & 13 & 12 & 11.9 & 44 & 02 & 16 & .S..6*/ & -15.43 & 4.8 & 407 & 179 & 3.24 & 0 \\
8303 &  & 13 & 13 & 17.6 & 36 & 12 & 58 & .IXS9.. & -17.07 & 17.2 & 946 & 89 & 9.97 & 0 \\
8320 & IC 859 & 13 & 14 & 26.4 & 45 & 55 & 34 & .IB.9.. & -15.35 & 5.9 & 195 & 60 & 5.93 & 0 \\
8331 & DDO 169 & 13 & 15 & 30.1 & 47 & 29 & 60 & .IA.9.. & -13.95 & 5.9 & 260 & 50 & 1.21 & 0 \\
8441 & DDO 175 & 13 & 25 & 29.1 & 57 & 49 & 20 & .I..9.. & -16.7 & 20.0 & 1519 & 103 & 15 & 0 \\
8489 & DDO 176 & 13 & 29 & 38.8 & 45 & 23 & 17 & .SX.8.. & -16.7 & 20.0 & 1303 & 130 & 12.5 & 0 \\
8490 & NGC 5204 & 13 & 29 & 36.4 & 58 & 25 & 12 & .SAS9.. & -16.8 & 4.9 & 204 & 110 & 5.97 & 0.01 \\
8508 & I Zw 60 & 13 & 30 & 44.5 & 54 & 54 & 39 & .IA.9.. & -13.43 & 3.7 & 62 & 49 & 0.5 & 0 \\
8550 & NGC 5229 & 13 & 34 & 03.0 & 47 & 54 & 53 & .SBS7?/ & -14.31 & 5.3 & 364 & 127 & 1.79 & 0 \\
8565 & NGC 5238, I Zw 64 & 13 & 34 & 42.8 & 51 & 36 & 52 & .SXS8.. & -14.67 & 5.2 & 232 & 36 & 0.42 & 0 \\
8651 & DDO 181 & 13 & 39 & 53.9 & 40 & 44 & 21 & .I..9.. & -14.15 & 5.9 & 203 & 41 & 1.06 & 0 \\
8683 & DDO 182 & 13 & 42 & 33.1 & 39 & 39 & 28 & .I..9.. & -15.29 & 12.6 & 661 & 35 & 2.68 & 0 \\
8760 & DDO 183 & 13 & 50 & 50.5 & 38 & 01 & 5 & .I..9.. & -13.95 & 5.9 & 193 & 32 & 0.83 & 0 \\
8837 & DDO 185 & 13 & 54 & 45.8 & 53 & 54 & 10 & .IBS9./ & -14.83 & 5.1 & 144 & 77 & 1.24 & 0 \\
8892 &  & 13 & 57 & 41.1 & 57 & 00 & 6 & .I..9.. & -17.3 & 29.0 & 1748 & 95 & 24 & 0 \\
9018 & NGC 5477, DDO 186 & 14 & 05 & 32.6 & 54 & 27 & 39 & .SAS9.. & -15.22 & 7.7 & 304 & 52 & 1.84 & 0 \\
9128 & DDO 187 & 14 & 15 & 56.7 & 23 & 03 & 18 & .I..9.. & -13.91 & 4.4 & 154 & 33 & 0.58 & 0 \\
9211 & DDO 189 & 14 & 22 & 32.3 & 45 & 22 & 60 & .I..9*. & -15.49 & 12.6 & 690 & 99 & 9.21 & 0 \\
9219 & NGC 5608 & 14 & 23 & 17.6 & 41 & 46 & 31 & .I..9*. & -16.49 & 12.0 & 663 & 109 & 4.19 & 0 \\
9240 & I Zw 87 & 14 & 24 & 43.6 & 44 & 31 & 37 & .IA.9.. & -14.66 & 4.5 & 153 & 45 & 1.14 & 0 \\
9405 & DDO 194 & 14 & 35 & 24.2 & 57 & 15 & 19 & .I..9.. & -- & 7.6 & 222 & 85 & 1.38 & 0 \\
9426 &  & 14 & 37 & 29.1 & 48 & 37 & 26 & .I..9.. & -- & 36.2 & 2311 & 44 & 14.2 & 0.08 \\
9769 &  & 15 & 12 & 07.2 & 55 & 47 & 6 & .SXT8*. & -16.29 & 16.6 & 844 & 129 & 14.7 & 0.01 \\
9906 & NGC 5963 & 15 & 33 & 27.8 & 56 & 33 & 36 & .S...P. & -17.97 & 13.7 & 656 & 214 & 20.7 & 0 \\
9992 &  & 15 & 41 & 48.0 & 67 & 15 & 15 & .I..9.. & -14.88 & 10.4 & 427 & 48 & 2.62 & 0.05 \\
10736 &  & 17 & 08 & 04.6 & 69 & 27 & 53 & .SX.8.. & -15.83 & 11.7 & 490 & 144 & 7.31 & 0.1 \\
10792 &  & 17 & 14 & 02.3 & 75 & 12 & 13 & .I..9.. & -- & 22.2 & 1233 & 62 & 9.13 & 0.09 \\
10806 &  & 17 & 18 & 51.1 & 49 & 53 & 2 & .SBS8.. & -17.5 & 17.7 & 930 & 152 & 16.7 & 0.04 \\
11283 & IC 1291 & 18 & 33 & 52.6 & 49 & 16 & 44 & .SBS8?. & -18.57 & 31.3 & 1962 & 173 & 48.1 & 0.15 \\
11557 &  & 20 & 24 & 00.7 & 60 & 11 & 41 & .SXS8.. & -18.17 & 23.8 & 1389 & 87 & 26.5 & 0.65 \\
11707 &  & 21 & 14 & 31.7 & 26 & 44 & 5 & .SA.8.. & -16.5 & 15.9 & 906 & 185 & 34.2 & 0.36 \\
12048 & NGC 7292 & 22 & 28 & 25.8 & 30 & 17 & 34 & .IB.9.. & -18.18 & 16.8 & 986 & 77 & 13.3 & 0.14 \\
12060 &  & 22 & 30 & 33.9 & 33 & 49 & 14 & .IB.9.. & -16.07 & 15.7 & 884 & 113 & 14.9 & 0.2 \\
12212 &  & 22 & 50 & 30.2 & 29 & 08 & 20 & .S..9*. & -- & 15.5 & 894 & 105 & 8.88 & 0.14 \\
12554 & NGC 7640 & 23 & 22 & 06.7 & 40 & 50 & 43 & .SBS5.. & -18.35 & 9.2 & 369 & 234 & 65.9 & 0.24 \\
12632 & DDO 217 & 23 & 29 & 58.8 & 40 & 59 & 27 & .S..9*. & -14.69 & 6.9 & 422 & 114 & 8.19 & 0.28 \\
12732 &  & 23 & 40 & 39.8 & 26 & 14 & 11 & .S..9*. & -16.79 & 13.2 & 748 & 112 & 28.1 & 0.1 \\
\noalign{\smallskip}\hline
\end{tabular}
\end{flushleft}
\vfill
\end{table*}
\addtocounter{table}{-1}
\addtocounter{table}{1}
}

\newpage\clearpage

\begin{table}
\caption[]{Adopted distances}\label{disttable}
\null\vspace{-0.5cm}
\end{table}
\begin{figure}[t]
\vspace{-0.25cm}
\resizebox{\hsize}{!}{\includegraphics{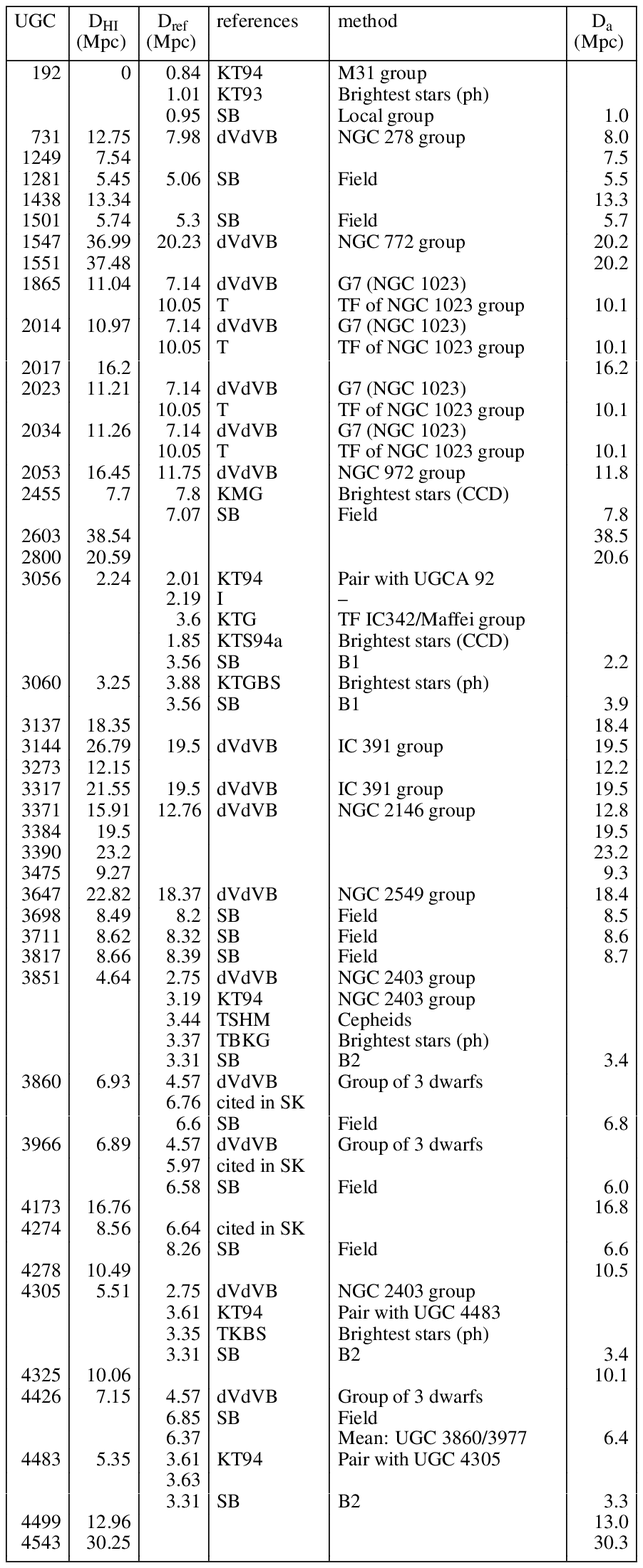}}
\end{figure}
\begin{figure}[t]
\vspace{0.4cm}
\resizebox{\hsize}{!}{\includegraphics{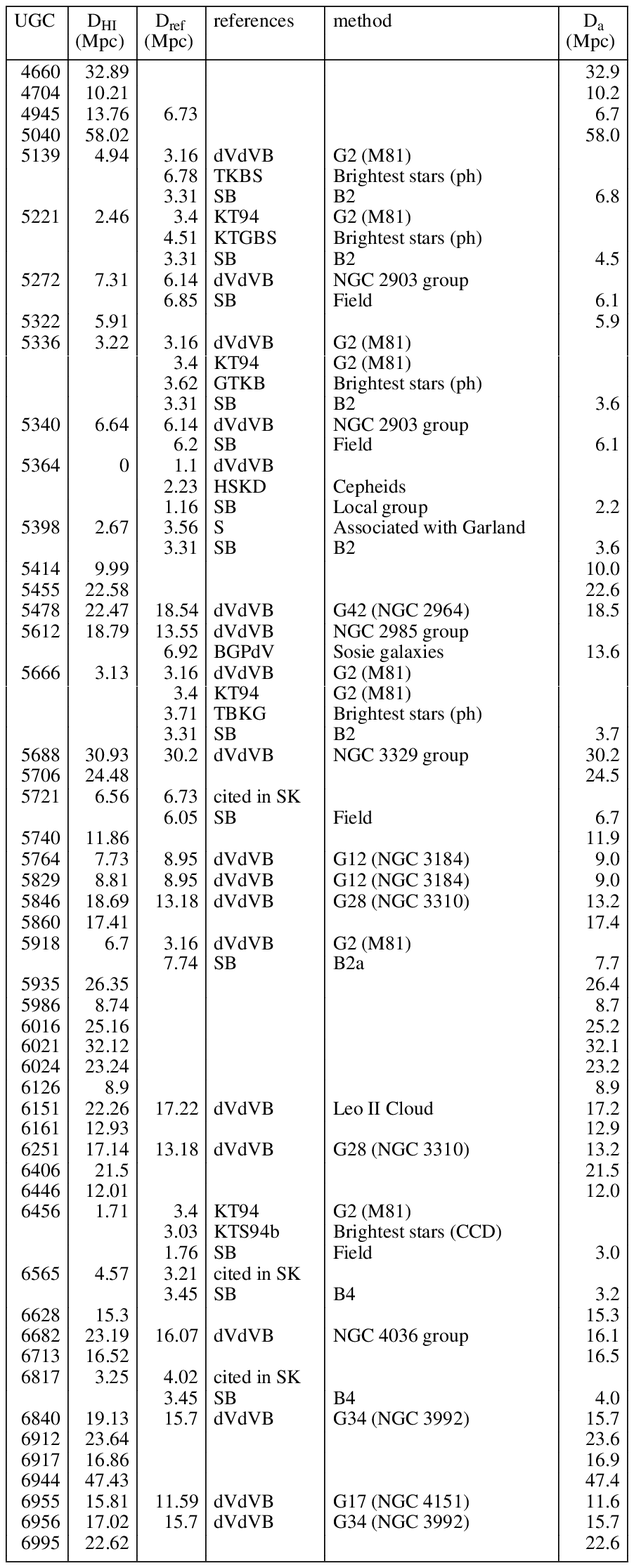}}
\end{figure}
\begin{figure}[t]
\resizebox{\hsize}{!}{\includegraphics{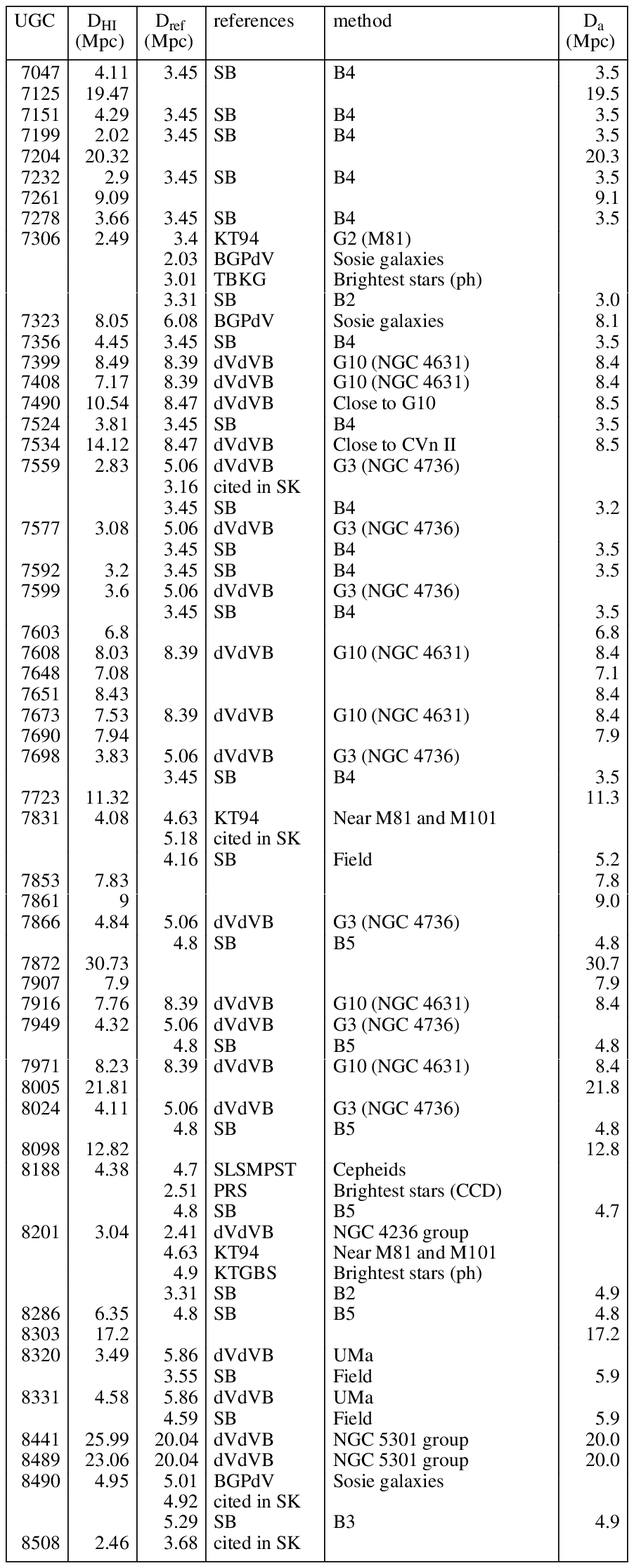}}
\end{figure}
\begin{figure}[t]
\resizebox{\hsize}{!}{\includegraphics{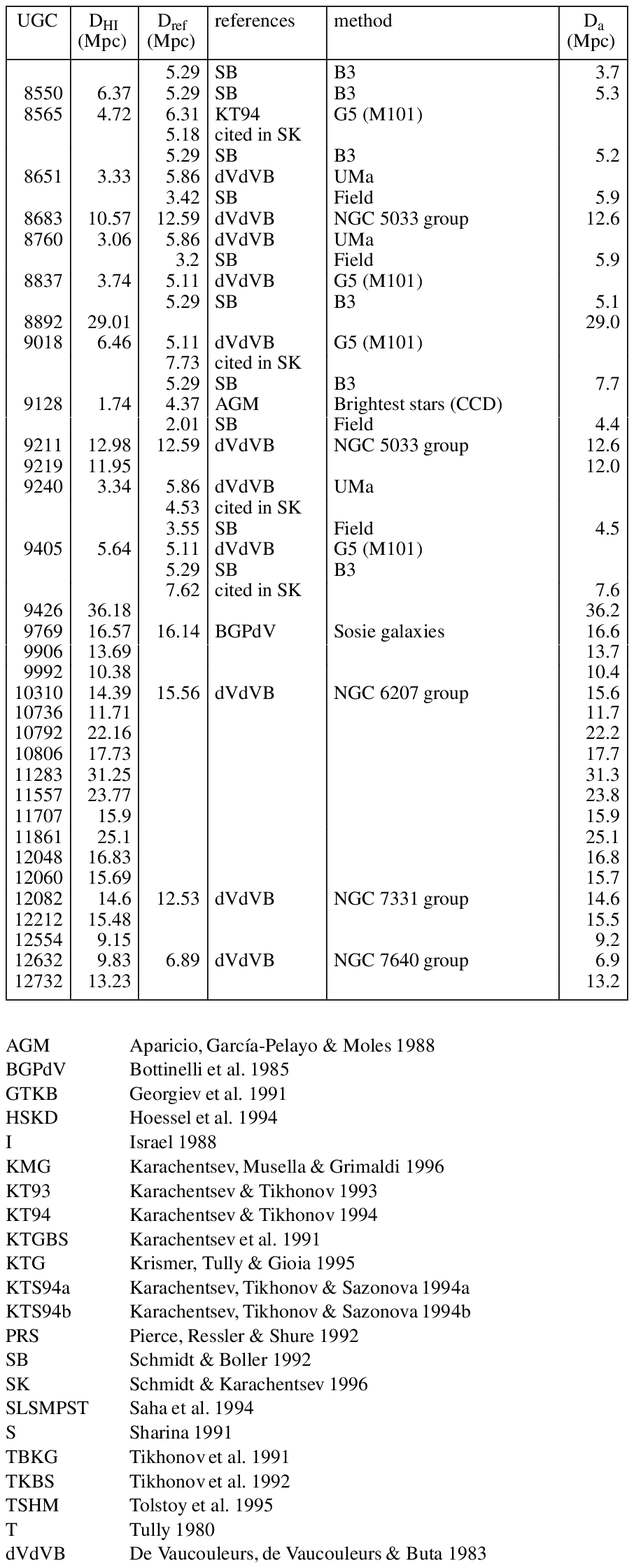}}
\end{figure}

\newpage\clearpage

\setlength{\tabcolsep}{6pt}
\begin{table}
\caption[]{List of $R$-band observations\\}\label{obslistR}
\begin{flushleft}
\begin{tabular}{rcrcr}
\hline\noalign{\smallskip}
UGC & date & $t_{\rm exp}$ & seeing & $\sigma_{\rm phot}$ \\
 &  & sec & $''$ & mag \\
(1) & (2) & (3) & (4) & (5) \\
\noalign{\smallskip}\hline\noalign{\smallskip}
192 & ~1994  Nov  29~ & 60 & \hbox to1.3cm{\hfil2.2\kern0.45cm} & 0.05\\
731 & ~1995  Dec  23~ & 600 & \hbox to1.3cm{\hfil1.0\kern0.45cm} & 0.02\\
1249 & ~1994  Dec  2~ & 600 & \hbox to1.3cm{\hfil1.6\kern0.45cm} & 0.22\\
1281 & ~1995  Dec  24~ & 300 & \hbox to1.3cm{\hfil0.9\kern0.45cm} & 0.02\\
1438 & ~1994  Dec  2~ & 600 & \hbox to1.3cm{\hfil1.1\kern0.45cm} & 0.22\\
\hline
\end{tabular}
\end{flushleft}
\end{table}
\addtocounter{table}{-1}
\begin{table}
\caption[]{-- Continued}
\begin{flushleft}
\begin{tabular}{rcrcr}
\hline\noalign{\smallskip}
UGC & date & $t_{\rm exp}$ & seeing & $\sigma_{\rm phot}$ \\
 &  & sec & $''$ & mag \\
(1) & (2) & (3) & (4) & (5) \\
\noalign{\smallskip}\hline\noalign{\smallskip}
5478 & ~1995  Dec  25~ & 600 & \hbox to1.3cm{\hfil1.7\kern0.45cm} & 0.07\\
5612 & ~1995  Feb  7~ & 600 & \hbox to1.3cm{\hfil1.4\kern0.45cm} & 0.51\\
5666 & ~1995  Dec  28~ & 360 & \hbox to1.3cm{\hfil1.8\kern0.45cm} & 0.19\\
5688 & ~1994  May  6~ & 600 & \hbox to1.3cm{\hfil2.2\kern0.45cm} & 0.06\\
5706 & ~1995  Dec  25~ & 900 & \hbox to1.3cm{\hfil1.7\kern0.45cm} & 0.07\\
\hline
\end{tabular}
\end{flushleft}
\end{table}
\addtocounter{table}{-1}
\begin{table}
\caption[]{-- Continued}
\begin{flushleft}
\begin{tabular}{rcrcr}
\hline\noalign{\smallskip}
UGC & date & $t_{\rm exp}$ & seeing & $\sigma_{\rm phot}$ \\
 &  & sec & $''$ & mag \\
(1) & (2) & (3) & (4) & (5) \\
\noalign{\smallskip}\hline\noalign{\smallskip}
7648 & ~1994  May  7~ & 60 & \hbox to1.3cm{\hfil2.0\kern0.45cm} & 0.06\\
7651 & ~1994  May  7~ & 60 & \hbox to1.3cm{\hfil2.0\kern0.45cm} & 0.06\\
7673 & ~1996  May  11~ & 600 & \hbox to1.3cm{\hfil1.5\kern0.45cm} & 0.02\\
7690 & ~1994  May  3~ & 600 & \hbox to1.3cm{\hfil1.9\kern0.45cm} & 0.01\\
7698 & ~1994  May  5~ & 1200 & \hbox to1.3cm{\hfil2.0\kern0.45cm} & 0.01\\
\hline
\end{tabular}
\end{flushleft}
\end{table}
\addtocounter{table}{-1}
\addtocounter{table}{1}

\setlength{\tabcolsep}{6pt}
\begin{table}
\caption[]{List of $B$-band observations\\}\label{obslistB}
\begin{flushleft}
\begin{tabular}{rcrcr}
\hline\noalign{\smallskip}
UGC & date & $t_{\rm exp}$ & seeing & $\sigma_{\rm phot}$ \\
 &  & s & $''$ & mag \\
(1) & (2) & (3) & (4) & (5) \\
\noalign{\smallskip}\hline\noalign{\smallskip}
731 & ~23 Dec 1995~ & 600 & \hbox to1.3cm{\hfil1.0\kern0.45cm} & 0.03\\
1281 & ~24 Dec 1995~ & 900 & \hbox to1.3cm{\hfil1.1\kern0.45cm} & 0.09\\
1501 & ~24 Dec 1995~ & 600 & \hbox to1.3cm{\hfil1.1\kern0.45cm} & 0.09\\
2800 & ~23 Dec 1995~ & 600 & \hbox to1.3cm{\hfil1.1\kern0.45cm} & 0.03\\
3060 & ~24 Dec 1995~ & 600 & \hbox to1.3cm{\hfil1.3\kern0.45cm} & 0.09\\
\end{tabular}
\end{flushleft}
\end{table}

}\newpage\clearpage

{\renewcommand{\baselinestretch}{0.99}

\setlength{\tabcolsep}{4.5pt}
\def\ruleit#1{\vrule width#1 height0pt depth0pt}
{
\begin{table*}
\null\vspace{1pt}
\caption[]{$R$-band isophotal and photometric parameters}\label{globpropR}
\begin{flushleft}
\begin{tabular}{rrrrrrrrrrrrrrrrr}
\hline\noalign{\smallskip}
UGC ~& $\epsilon$ & PA & $m_{25}$ & $m_{\rm lim}$ & $m_{\rm ext}$ & $\mu_{\rm lim}$ & $\mu_{\rm c}$ & $\mu_0$ & M$_R$ & $\sigma_{\rm phot}$ & $h$ & $d_{25}$ & $d_{26.5}$ & $r_{20}$ & $r_{50}$ & $r_{80}$ \\
\noalign{\smallskip}
 (1) & (2) & (3) & (4) & (5) & (6) & (7) & (8) & (9) & (10) & (11) & (12) & (13) & (14) & (15) & (16) & (17) \\
\noalign{\smallskip}\hline\noalign{\smallskip}
 192\rlap{$^a$} ~& 0.474 & 133 & 9.75 & 9.93 & 9.79 & 24.08 & 21.35 & 20.76 & -15.21 & 0.05 & 92.7 & 1015 & -- &   74 &  145 &  239 \\
 731 ~& 0.506 & 73 & 13.50 & 13.52 & 13.29 & 24.95 & 21.86 & 22.24 & -16.63 & 0.02 & 45.8 & 249 & -- &   29 &   63 &  104 \\
 1249\rlap{$^b$} ~& 0.629 & 151 & 11.63 & 11.60 & 11.58 & 25.78 & 22.13 & 21.05 & -17.94 & 0.22 & 56.4 & 396 & 463 &   42 &   91 &  144 \\
 1281 ~& 0.872 & 40 & 12.72 & 12.61 & 12.61 & 26.64 & 21.00 & 20.51 & -16.17 & 0.02 & 45.9 & 359 & 582 &   41 &   79 &  133 \\
 1438\rlap{$^b$} ~& 0.347 & 87 & 12.70 & 12.65 & 12.65 & 27.62 & 19.59 & 19.52 & -18.27 & 0.22 & 13.6 & 140 & 184 &   11 &   22 &   41 \\
 1501 ~& 0.774 & 1 & 11.70 & 11.65 & 11.63 & 26.10 & 21.28 & 20.65 & -17.27 & 0.02 & 58.5 & 448 & 592 &   47 &   93 &  158 \\
\hline
\end{tabular}
\end{flushleft}
\end{table*}
\addtocounter{table}{-1}
\clearpage
\begin{table*}
\null\vspace{1pt}
\caption[]{-- Continued}
\begin{flushleft}
\begin{tabular}{rrrrrrrrrrrrrrrrr}
\hline\noalign{\smallskip}
UGC ~& $\epsilon$ & PA & $m_{25}$ & $m_{\rm lim}$ & $m_{\rm ext}$ & $\mu_{\rm lim}$ & $\mu_{\rm c}$ & $\mu_0$ & M$_R$ & $\sigma_{\rm phot}$ & $h$ & $d_{25}$ & $d_{26.5}$ & $r_{20}$ & $r_{50}$ & $r_{80}$ \\
\noalign{\smallskip}
 (1) & (2) & (3) & (4) & (5) & (6) & (7) & (8) & (9) & (10) & (11) & (12) & (13) & (14) & (15) & (16) & (17) \\
\noalign{\smallskip}\hline\noalign{\smallskip}
 5340 ~& 0.501 & 8 & 14.57 & 14.44 & 14.37 & 26.07 & 21.85 & 21.99 & -14.57 & 0.01 & 18.9 & 103 & 169 &   13 &   27 &   48 \\
 5364 ~& 0.497 & 102 & 13.02 & 12.84 & 12.60 & 25.51 & 23.71 & 22.78 & -14.15 & 0.01 & 69.3 & 276 & 455 &   57 &  102 &  165 \\
 5398 ~& 0.231 & 49 & 9.51 & 9.51 & 9.47 & 25.19 & 17.96 & 19.89 & -18.43 & 0.01 & 51.8 & 478 & -- &   25 &   65 &  132 \\
 5414 ~& 0.416 & 38 & 12.59 & 12.48 & 12.47 & 27.82 & 21.85 & 21.18 & -17.55 & 0.02 & 30.4 & 210 & 270 &   27 &   53 &   85 \\
 5455 ~& 0.440 & 147 & 15.85 & 15.37 & 15.33 & 27.22 & 23.02 & 22.47 & -16.54 & 0.05 & 15.1 & 68 & 119 &   13 &   26 &   47 \\
 5478 ~& 0.000 & 42 & 15.19 & 15.04 & 14.57 & 25.42 & 22.62 & 22.91 & -16.80 & 0.07 & 18.7 & 70 & 115 &   15 &   30 &   58 \\
\hline
\end{tabular}
\end{flushleft}
\end{table*}
\addtocounter{table}{-1}
\clearpage
\begin{table*}
\null\vspace{1pt}
\caption[]{-- Continued}
\begin{flushleft}
\begin{tabular}{rrrrrrrrrrrrrrrrr}
\hline\noalign{\smallskip}
UGC ~& $\epsilon$ & PA & $m_{25}$ & $m_{\rm lim}$ & $m_{\rm ext}$ & $\mu_{\rm lim}$ & $\mu_{\rm c}$ & $\mu_0$ & M$_R$ & $\sigma_{\rm phot}$ & $h$ & $d_{25}$ & $d_{26.5}$ & $r_{20}$ & $r_{50}$ & $r_{80}$ \\
\noalign{\smallskip}
 (1) & (2) & (3) & (4) & (5) & (6) & (7) & (8) & (9) & (10) & (11) & (12) & (13) & (14) & (15) & (16) & (17) \\
\noalign{\smallskip}\hline\noalign{\smallskip}
 7408 ~& 0.279 & 96 & 13.22 & 13.08 & 13.02 & 25.99 & 21.60 & 21.56 & -16.60 & 0.01 & 24.2 & 155 & 234 &   20 &   40 &   71 \\
 7490 ~& 0.000 & 90 & 12.44 & 12.31 & 12.25 & 26.54 & 21.31 & 21.33 & -17.40 & 0.01 & 27.3 & 180 & 257 &   24 &   47 &   81 \\
 7524\rlap{$^b$} ~& 0.368 & 131 & 9.76 & 9.65 & 9.58 & 26.04 & 20.97 & 21.70 & -18.14 & 0.18 & 135.1 & 867 & 1111 &  116 &  225 &  395 \\
 7534 ~& 0.000 & 36 & 13.84 & 13.75 & 13.46 & 25.59 & 0.00 & 22.45 & -16.19 & 0.01 & 24.8 & 120 & 167 &   25 &   42 &   76 \\
 7559\rlap{$^b$} ~& 0.476 & 157 & 14.14 & 13.87 & 13.87 & 25.69 & 22.80 & 23.08 & -13.66 & 0.06 & 44.6 & 157 & 283 &   26 &   51 &   81 \\
 7577 ~& 0.577 & 127 & 12.25 & 12.14 & 12.10 & 26.08 & 22.28 & 21.56 & -15.62 & 0.02 & 50.8 & 318 & 451 &   43 &   82 &  137 \\
\hline
\end{tabular}
\end{flushleft}
\end{table*}
\addtocounter{table}{-1}
\clearpage
\begin{table*}
\null\vspace{1pt}
\caption[]{-- Continued}
\begin{flushleft}
\begin{tabular}{rrrrrrrrrrrrrrrrr}
\hline\noalign{\smallskip}
UGC ~& $\epsilon$ & PA & $m_{25}$ & $m_{\rm lim}$ & $m_{\rm ext}$ & $\mu_{\rm lim}$ & $\mu_{\rm c}$ & $\mu_0$ & M$_R$ & $\sigma_{\rm phot}$ & $h$ & $d_{25}$ & $d_{26.5}$ & $r_{20}$ & $r_{50}$ & $r_{80}$ \\
\noalign{\smallskip}
 (1) & (2) & (3) & (4) & (5) & (6) & (7) & (8) & (9) & (10) & (11) & (12) & (13) & (14) & (15) & (16) & (17) \\
\noalign{\smallskip}\hline\noalign{\smallskip}
 9769 ~& 0.353 & 9 & 13.24 & 13.10 & 13.01 & 26.38 & 20.69 & 22.24 & -18.10 & 0.01 & 33.5 & 170 & 238 &   18 &   44 &   83 \\
 9906 ~& 0.307 & 51 & 11.98 & 11.86 & 11.81 & 26.08 & 18.73 & 22.73 & -18.87 & 0.01 & 45.5 & 179 & 310 &    9 &   19 &   66 \\
 9992 ~& 0.319 & 148 & 14.53 & 14.26 & 14.24 & 27.65 & 21.72 & 21.75 & -15.89 & 0.04 & 15.9 & 90 & 149 &   13 &   26 &   49 \\
 10310 ~& 0.298 & 108 & 13.24 & 13.11 & 13.07 & 26.62 & 21.55 & 21.66 & -17.90 & 0.02 & 24.9 & 158 & 217 &   20 &   42 &   71 \\
 10736 ~& 0.727 & 156 & 13.81 & 13.75 & 13.73 & 25.95 & 21.11 & 21.13 & -16.71 & 0.02 & 27.1 & 183 & 229 &   21 &   43 &   68 \\
 10792 ~& 0.339 & 94 & 16.07 & 16.00 & 15.77 & 25.26 & 22.78 & 22.27 & -16.05 & 0.02 & 10.9 & 54 & -- &    9 &   17 &   29 \\
\end{tabular}
\end{flushleft}
\end{table*}
}

\newpage\clearpage

\setlength{\tabcolsep}{4.5pt}
\def\ruleit#1{\vrule width#1 height0pt depth0pt}
{
\begin{table*}
\null\vspace{1pt}
\caption[]{$B$-band isophotal and photometric parameters\\}\label{globpropB}
\begin{flushleft}
\begin{tabular}{rrrrrrrrrrrrrrrrr}
\hline\noalign{\smallskip}
UGC ~& $\epsilon$ & PA & $m_{25}$ & $m_{\rm lim}$ & $m_{\rm ext}$ & $\mu_{\rm lim}$ & $\mu_{\rm c}$ & $\mu_0$ & M$_R$ & $\sigma_{\rm phot}$ & $h$ & $d_{25}$ & $d_{26.5}$ & $r_{20}$ & $r_{50}$ & $r_{80}$ \\
\noalign{\smallskip}
 (1) & (2) & (3) & (4) & (5) & (6) & (7) & (8) & (9) & (10) & (11) & (12) & (13) & (14) & (15) & (16) & (17) \\
\noalign{\smallskip}\hline\noalign{\smallskip}
 731 ~& 0.506 & 73 & 15.12 & 14.43 & 14.14 & 25.78 & 22.98 & 23.27 & -15.78 & 0.03 & 50.5 & 154 & 284 &   34 &   70 &  110 \\
 1281 ~& 0.872 & 40 & 13.63 & 13.50 & 13.49 & 26.92 & 21.74 & 21.10 & -15.29 & 0.09 & 41.9 & 290 & 405 &   39 &   75 &  121 \\
 1501 ~& 0.774 & 1 & 12.49 & 12.37 & 12.36 & 27.19 & 21.99 & 21.29 & -16.54 & 0.09 & 55.8 & 376 & 497 &   45 &   90 &  151 \\
 2800 ~& 0.604 & 106 & 15.61 & 15.08 & 15.06 & 28.35 & 21.97 & 22.16 & -17.28 & 0.03 & 25.5 & 137 & 194 &   23 &   43 &   70 \\
 3060 ~& 0.789 & 24 & 12.19 & 11.98 & 11.97 & 27.96 & 21.76 & 21.45 & -16.34 & 0.09 & 81.3 & 540 & 773 &   65 &  135 &  237 \\
 3137\rlap{$^b$} ~& 0.765 & 74 & 15.1 & 14.7 & 14.7 & 27.0 & 22.7 & 22.1 & -16.9 & 0.45 & 29.4 & 156 & 240 &   25 &   47 &   82 \\
\end{tabular}
\end{flushleft}
\end{table*}
}

\newpage\clearpage
}

\section{Images and surface brightness profiles}
\label{thefigures}

The next pages show an overview of the images, surface brightness
profiles and isophotal fits for the 171 late-type dwarf galaxies in
the sample. For each galaxy one row with three panels is shown.

The first panel gives the $R$-band image. All images have been put on
a common grayscale using the calibration described in
Sect.~\ref{observations}, and without correcting for Galactic
foreground extinction. On the top left of each panel the UGC number is
given, along with at most one other common name (either NGC or
DDO). In the lower left a yardstick is shown, where the size depends
on the scale of the galaxy. This size is shown above the yardstick and
represents 1, 2, 5 or 10 kpc. The white cross near the center of the
galaxy gives the adopted center from the isophotal fits (see
Sect.~\ref{profiles}). The ellipse indicates the orientation
parameters, derived from the isophotal fits, that were adopted for
each galaxy and that were used to derive the radial surface brightness
profile.

The second panel presents the radial surface brightness profile. On
the top right the morphological type, the absolute $R$-band magnitude
and the conversion between arcseconds and parsecs are given. The arrow
on the left side indicates the $3\sigma$ above sky level.  The radial
surface brightness profile has not been corrected for Galactic
foreground extinction. The radial scale is measured along the major
axis.

The third panel gives the results from the isophotal fits with all
parameters (center, position angle and ellipticity) free. Note that
these are not the data from which the final orientation parameters
were derived. These were derived in several steps as described in
Sect.~\ref{profiles}. Within this panel there are four subpanels.
Clockwise from the top left these show the variation of ellipticity,
position angle, $y$-position and $x$-position of the center with
radius. The dotted line in each panel gives the adopted value
for the plotted parameter.

\vspace{1cm}

\noindent {\bf Full version with figures in this appendix can be
downloaded from:

\noindent http://www.robswork.net/publications/WHISPII.ps.gz}

\end{document}